\documentclass[12pt]{aastex62}

\begin{document}

\parindent=1.5cm

\title{MORE OLD DOGS, MORE NEW TRICKS: PHOTOGRAPHIC PLATES 
OF THE GLOBULAR CLUSTER M10 (NGC 6254) DIGITIZED WITH A COMMERCIAL SCANNER}

\author{T. J. Davidge}
\affiliation{Dominion Astrophysical Observatory,
\\Herzberg Astronomy \& Astrophysics Research Center,
\\National Research Council of Canada, 5071 West Saanich Road,
\\Victoria, BC Canada V9E 2E7\\tim.davidge@nrc.ca; tdavidge1450@gmail.com}

\begin{abstract}

	Photographic plates of the globular cluster M10 (NGC 6254) 
that were recorded at the Dominion Astrophysical Observatory 
between 1931 and 1934 have been digitized with a commercial scanner. 
The plates have numerous cosmetic issues, and the results are used to assess 
the information that can be extracted from such archival data. After 
performing a series of basic processing steps, a plate-to-plate dispersion 
in photometric measurements of $\leq 10\%$ is delivered between 
$V =$ 13 and 15. The light curves of known variables are constructed and 
evidence is presented for long term variations in the mean brightnesses 
of the W Vir star V3, as well as the semi-regular (SR) variable V29. 
Two new variable stars are identified, although cluster membership 
is possible for only one of these. A brightness-selected sample of candidate 
variable stars identified in the GAIA database that are not in the 
General Catalogue of Variable Stars is also examined, and some 
show variability in the plate photometry. Parallaxes indicate that five of 
the GAIA variables are at the same distance as M10. Four of the GAIA 
variables fall along the horizontal branch of M10 on the color-magnitude 
diagram. Another GAIA variable is located close to the cluster center, and 
may be evolving on the supra-horizontal branch. The plate photometry 
suggests that the Gaia variable with the most secure M10 
membership based on parallax, proper motions, and 
velocity is an SR variable that is offset by 
a projected distance of 9 parsecs from the cluster center. 

\end{abstract}

\section{INTRODUCTION}

	Photographic plates were the only wide-field detector available 
for imaging studies throughout much of the last century. The collections 
of photographic plates obtained for targets such as star clusters, 
the Magellanic Clouds, and other fields of interest, coupled with 
the extensive photographic all-sky surveys that were obtained during the 
pre-CCD era, form datasets that span a number of decades. With the exception 
of visual and photoelectric measurements of single objects, plate collections 
thus provide a key means of tracing the properties of variable stars prior 
to the wide-spread use of digital detectors.

	A potential problem with plate collections, especially those recorded 
for personal use and/or of a specific target, is that they can 
form diverse datasets, due to differences in observing 
conditions, the evolution of emulsion types with time, 
cosmetic issues related to their development and handling, and the inevitable 
aging of the plates. In addition, many photographic plates 
have not been digitized, despite the wealth of information that they contain 
\citep[e.g.][]{whietal2024}. In recent years there have been efforts 
to digitize plate libraries, such as those at Harvard Observatory
\citep[][]{grietal2009}, using specialized 
equipment \citep[][]{simetal2006}. While this is certainly a forefront 
endeavor, the use of a specialized scanning device like that 
described by \citet{simetal2006} may not be a practical option for astronomers 
who have undigitized plates at their home institutions.

	Commercial scanners are a practical alternative for digitizing plates. 
Top-of-the-line commercial scanners, such as the Epson 12000XL, require only a 
modest capital and operational investment when compared with more specialized 
equipment. The ability of the 12000XL to recover information from 
photographic spectra was examined by \citet{dav2024,dav2025}. 
\citet{ceretal2021} and \citet{aoketal2021} used similar devices 
to digitize plates in their collections. Once the plates are digitized, it 
becomes possible to place quantifiable limits on the information 
that can be extracted. Here, we examine imaging plates recorded at the Dominion 
Astrophysical Observatory (DAO) in the first half of the last century 
as part of a survey of globular clusters. 

	A number of photographic plates of the globular cluster M10 
(NGC 6254) were recorded at the Newtonian focus of the 1.8 meter telescope 
at the DAO during the early 1930s. These plates were part of a program 
discussed by \citet{saw1938a, saw1938b} to search for variable stars in 
globular clusters. There are plates of a number of clusters in the DAO 
archive, and M10 was selected for the present study because it is 
at a distance of a few kpc, and so the variable stars are expected to be 
bright when compared with those in other clusters. As will be shown 
in Section 2, the plates have a highly heterogeneous 
appearance - they constitute what is likely a 'typical' collection 
of plates from this epoch that were recorded for single purpose use. There is 
thus an obvious interest in assessing the information that can be extracted 
from this material.

	M10 is projected against the Galactic Bulge at a galactic latitude 
of 23 degrees, and there is a rich population of 
field stars that are significant contaminants 
outside of the dense central regions of the cluster. Basic parameters of M10, 
taken from the Catalogue of Parameters for Milky-Way Globular Clusters 
\citep[][]{har1996}, are listed in Table 1. In Section 5.1 a mean 
parallax for M10 that is based on measurements in the GAIA DR3 
\citep[][]{gai2023} and that places M10 at a slightly greater distance than 
that in Table 1 is calculated.

\begin{deluxetable}{lc}
\tablecaption{M10 Parameters}
\tablehead{Parameter & Value \\}
\startdata
Distance & 4.4 kpc \\
M$_V$ & --7.5 \\
Fe/H & --1.56 \\
Core radius & 0.77 pc \\
Half light radius & 1.95 pc \\
Heliocentric radial velocity & 75.6 km/sec \\
\enddata
\end{deluxetable}

	\citet{saw1938a} identified variable stars by visually searching 
for objects that changed brightness from plate-to-plate. This technique has 
the potential to detect large amplitude variations among moderately bright 
stars, but may miss smaller amplitude variations and/or fainter objects. 
\citet{saw1938a} found two variables (V1 and V2) with this technique, while 
\citet{arp1955} found a third (V3) from other photographic observations. 
V1 is a semi-regular (SR) variable, while V2 and V3 are W Vir stars.

	\citet{karetal2022} examined the long-term behaviour of V1, V2, and V3. 
using information from photographic and CCD observations. 
While they considered material that spanned roughly a century, that study 
relied on the original published times of maximum and minimum 
light that were obtained directly from the older photographic material, 
rather than a re-assessment of the light curves based on digitized material. 
This resulted in substantial uncertainties when assessing the period of 
V3. The digitization of the plates in the DAO collection opens the 
possibility of better characterizing the light curves of previously identified 
variables almost a century in the past with the potential bonus of 
discovering variables that may have eluded detection.

	Numerous other variables have since been discovered with CCD 
observations in and around M10, and these include a number of SX Phe stars 
and SR variables \citep[][]{vonetal2002,saletal2016,rozetal2018,feretal2020}.
Many of the SR variables have observations that sample only a fraction of 
their light curves \citep[e.g. Figure 3 of][]{feretal2020} due to the 
long timescales of their variations. This highlights the importance of 
obtaining additional observations of these stars to better characterise their 
photometric variations. 

	The current paper has three broad goals. The first is to assess the 
photometry that can be extracted from the digitized plates, with 
emphasis on the plate-to-plate consistency that 
is a core consideration when assessing stellar variability. The second is 
to examine the photometric properties of known variables, 
and determine if there has been changes in observables such as period, 
the amplitude of variations, and mean brightness. Finally, the properties of 
new variables, with emphasis on objects flagged as photometric 
variables in the GAIA DR3 archive, are also examined.

\section{THE PLATES}

	The plates used in this study were recorded 
at the f/5 Newtonian focus of the DAO 1.8 meter telescope 
over the course of 16 nights between September 1931 and August 1934. 
The angular sampling is 0.00446 cm/arcsec, and so each $10 \times 
10$ cm plate covers roughly 37 arcmin on a side. Nineteen of the 
plates of M10 that were obtained by \citet{saw1938a} are in the 
DAO plate collection. One of these is annotated with star location 
information, and so is not considered further.

	Table 2 provides details of the plates with comments about their 
cosmetic quality. The fourth column contains comments in the observing 
log for that observation; in most cases there are no 
special comments. Plate numbers were assigned at the time of 
observation. It is not known if or how the plates were hyper-sensitized. 
The plates were recorded over a range of lunar phases, 
with some showing obvious scattered moon light. 
The plates recorded in 1934 have the best image quality, and are free of 
scattered light. The last column provides notes about the overall 
appearance of the scanned plate. A 'stained' entry refers to cosmetic defects 
that are likely due to development and handling issues. We suspect 
that the plate recorded on August 24, 1932 was partially obscured 
by the plate cassette dark slide. 

\begin{deluxetable}{lccll}
\tablecaption{M10 Plates in the DAO Collection}
\tablehead{Date & Plate \# & JD & Observing Log & Other Notes \\
 & & -2420000 & Entry & \\}
\startdata
Sep 21, 1931 & 19970 & 6607.712 & '1918 emulsion' & Blue emulsion \\
July 26, 1932 & 20543 & 6915.796 & & \\
Aug 1, 1932 & 20556 & 6921.742 & & Stained, fogged \\
Aug 3, 1932 & 20572 & 6923.804 & & Stained \\
Aug 4, 1932 & 20583 & 6924.739 & & Fogged \\
Aug 24, 1932 & 20644 & 6944.785 & & \\
Aug 24, 1932 & 20645 & 6944.744 & & Partial obstruction \\
Aug 26, 1932 & 20673 & 6946.713 & & Stained \\
Aug 26, 1932 & 20674 & 6946.722 & & Stained \\
July 20, 1933 & 21398 & 7274.752 & 'Sky pretty bright' & Scattered light \\
July 21, 1933 & 21411 & 7275.751 & & Scattered light \\
Aug 21, 1933 & 21514 & 7306.756 & & Blue emulsion \\
Aug 22, 1933 & 21536 & 7307.754 & & Mottled \\
Aug 23, 1933 & 21552 & 7308.712 & 'Mirror figure bad' & Scattered light \\
Aug 24, 1933 & 21569 & 7309.695 & & Scattered light \\
Aug 8, 1934 & 23236 & 7658.724 & & \\
Aug 9, 1934 & 23253 & 7659.776 & 'Guide star faint' & \\
Aug 14, 1934 & 23307 & 7664.783 & & \\
\enddata
\end{deluxetable}

	In addition to the issues described above, 
some of the plates had a thin brown coating that was removed with 
the careful application of an opticians chamois. Grime of this nature 
is a source of variations in the photometric zeropoint 
across the plates, and residual amounts remained near the plate edges. 
The nicotine hue of this material and its oily texture suggests that 
the plates may have been inspected by someone who was smoking.

	Sixteen of the plates have a type 'B' emulsion, for which the 
peak response is between 0.54 and $0.64\mu$m \citep{mee1931}, and so 
overlaps with the wavelength coverage of the $V$ filter. 
The plates recorded on September 21, 1931 and August 21, 1933 
used a different emulsion, and the observing logs 
for the former make reference to a '1918 emulsion'. The 
response of these plates indicates that they have a bluer 
wavelength coverage than the others, and we suspect that 
they are Eastman 40 ('E40') plates, for which the peak response is between 
0.3 and $0.5\mu$m \citep{mee1931}.

	Each plate was exposed for 10 minutes, and stellar profiles are
typically 4.5 -- 5 arcsec full width at half maximum (FWHM). 
In some cases the stellar images are trailed, suggesting problems with 
guiding that are likely due to the high airmass 
of M10 when observed from the DAO. Additional details regarding the plates 
and the observations can be found in \citet{saw1938a}.

	Examples of plate quality are shown in Figure 1. 
The top row shows plates that are cosmetically 
good and that were recorded with the B (left hand) and E40 
(right hand) emulsions. The bottom row shows plates with 
cosmetic defects due to scattered light from the moon (left hand panel), and 
to problems during development and subsequent handling 
(right hand panel). The plates in the bottom row are extreme examples of 
these problems.

\begin{figure}
\figurenum{1}
\epsscale{1.0}
\plotone{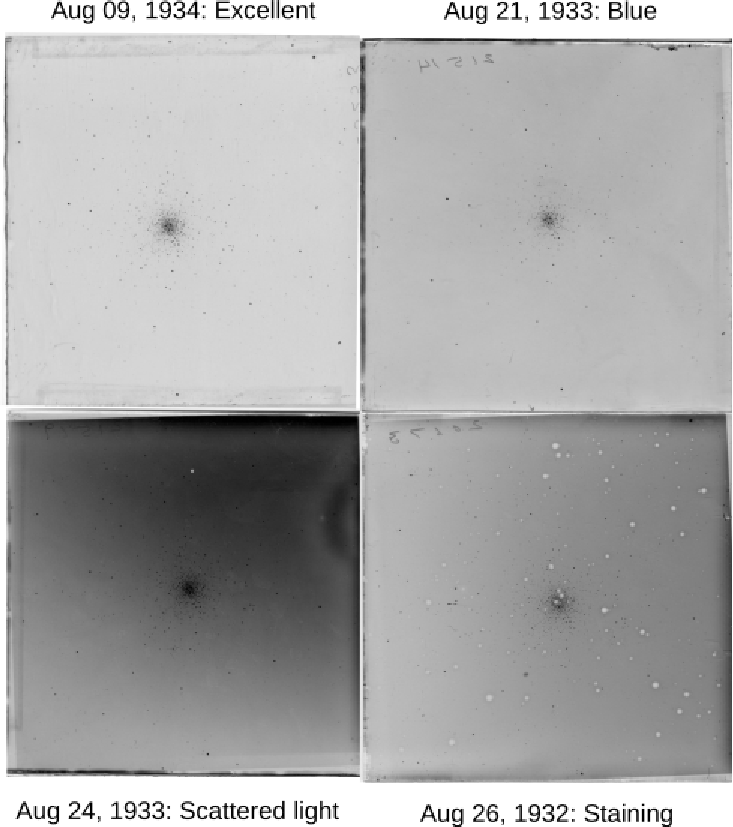}
\caption{Examples of plate quality. Starting from the upper left hand corner 
and moving clockwise are plates with (1) the B emulsion and excellent 
cosmetics (\# 23253), (2) the E40 emulsion and excellent cosmetics 
(\# 21514), (3) cosmetic defects that are likely due to problems 
with development, fixing, and/or handling (\# 20673), and (4) scattered 
moon light (\# 21569). Each plate covers $37 \times 37$ arcmin, 
with East to the right, and North at the bottom. (The FITS files of the 
unprocessed images shown in this figure, as well as those for the 
other nights, can be accessed in the online version of this paper).}
\end{figure}

\section{DIGITIZATION \& PROCESSING}

	The plates were digitized with the Epson 12000XL scanner 
that was used previously by \citet{dav2024, dav2025} to digitize spectra. 
The scan density was set at 1200 dpi to sample 
the FWHM of stars on the plate. Scanning was done 
in greyscale mode with 16 byte sampling.

	The images were initially saved as TIFF 
files and then converted into FITS format for subsequent processing.
The main processing goals are (1) to balance 
plate-to-plate consistency in photometric response, and (2) to construct a 
high S/N master reference image that is free of contamination from 
spurious objects such as dust spots and other cosmetic flaws. The frame 
constructed in the latter step allows a source catalogue to be obtained that is 
free of the cosmetic defects that can plague photographic plates. 
The challenges to constructing such a reference frame are evident in 
the lower row of Figure 1. While an alternative would be to use only the 
highest quality plates when making the reference image, there are only a 
handful of these, complicating the task of suppressing contaminants.

	Scattered light was suppressed by applying a 
boxcar median filter to each image, and then subtracting the result 
from the scanned image. The size of the filter was set so as to 
preserve light in the wings of the PSF. This filtering also suppresses
light from unresolved stars in the main body of the cluster. 

	Each plate has a unique angular scale that is defined by subtle 
differences in the positioning of the plate holder 
with respect to the telescope focus, coupled with thermal 
and gravity-induced variations in the telescope 
structure. Small scale distortions may also be introduced during scanning  
\citep[e.g.][]{aoketal2021}. The angular scale differences in the M10 
images amount to a few arcsec across each plate. The images were 
transformed to a common spatial scale and orientation by applying 
geometric corrections with the IRAF \citep[][]{tod1986, tod1993}
{\it geomap} and {\it geotran} tasks. The Aug 9, 1934 plate was selected as 
the reference for this step as it has good image quality and throughput.

	The next step was to compensate for plate-to-plate 
differences in the photometric zeropoint that may arise due to differences 
in sky transparency, emulsion properties, and varying amounts of scattered 
light. This was done by scaling the image intensities using aperture 
photometry of isolated stars that are moderately bright and 
unsaturated. This initial scaling was fine-tuned 
using photometric measurements made from PSF-fitting (next section). 

	The processed images were then smoothed with a Gaussian to produce 
plate-to-plate uniformity in angular resolution. While 
not necessary for photometric measurements 
in which the PSF is generated for each image (as is done 
here - see next section), this smoothing allowed a reference image to 
be constructed by stacking the processed images and finding the median 
intensity at each pixel location. The result is a 
high S/N image that is free of cosmetic issues, such as 
scratches, dust specs, pin holes in the emulsion, writing, etc. A catalogue 
of objects was obtained from the reference image, and this served as 
the basis for obtaining an initial set of photometric measurements. 
While the use of a single reference image to obtain a star catalogue 
produces a clean, uniform list of objects for photometry, 
transient objects, such as eruptive variables and cataclysmic variables 
will likely be missed, as they may not appear in the majority of images.

\section{PHOTOMETRIC MEASUREMENTS}

\subsection{Photometry and adjusting for zeropoint variations}

	Photometric measurements were made with the PSF-fitting 
routine ALLSTAR \citep[][]{steandhar1988} as implemented in IRAF. 
A PSF was constructed for each image from a fixed set of stars that are 
common to all images using the DAOPHOT \citep[][]{ste1987} PSF task.
PSF-fitting is an effective means of measuring brightnesses in 
crowded environments, such as globular clusters. 
However, a source of uncertainty in photometry obtained from photographic 
plates is the non-linear behaviour of photographic emulsions. The response is 
tracked by the characteristic curve, which can change from plate-to-plate 
due to differences in batch properties, age, and procedures used to increase 
sensitivity and tune linearity. One result is 
that the PSF may change shape with source brightness, 
depending on the location of the PSF core and wings on the 
characteristic curve \citep{ste1979}. 

	Two steps were taken to mitigate against the impact of non-linearity 
on PSF shape. First, PSF stars were selected to be in a brightness range where 
the response appears to be linear (see Figure 2, below). Second, a 
large fitting radius that extends well into the PSF wings 
was adopted. The shape of a typical characteristic curve is such that 
the signal in the PSF wings of a moderately bright star has a greater 
chance of falling in the linear response regime than the core. 
The use of a large fitting radius also allows the brightnesses of objects 
that are saturated, or are close to saturation, in their central regions to 
be more reliably measured.

	A sample of moderately bright stars in uncrowded fields that have 
modest plate-to-plate dispersions in their magnitudes 
were identified, and the brightnesses of these were used to compute 
refined nightly offsets to obtain improved plate-to-plate 
agreement in the photometric zeropoint. However, it became 
apparent while calculating these corrections that there were 
variations in the zeropoint across the images. These variations 
were typically on the order of 10\% or lower. 

	The non-uniformity in zeropoints was greatest among the plates 
recorded in 1933, with the intraplate zeropoint variations showing similar 
night-to-night behavior. It is unlikely that these variations are due to 
large-scale cosmetic defects across the plates 
as this would require fortuitous systematic differences 
involving the entire batch of plates that were used during that 
observing season, and/or the manner with which they 
were mounted on the telescope. A more likely explanation might be vignetting, 
perhaps by a partial obscuration that was fixed in place in the 
converging beam during the 1933 imaging run.

	Zeropoint variations across each plate were mapped using 
photometry of moderately bright and isolated stars that are 
distributed across the plates. A low order interpolation of the results 
across the images was then used to obtain a uniform zeropoint across 
each plate. This was done in such a way that the plate-to-plate 
uniformity in the mean zeropoint was also maintained. This 
procedure did not track variations on small angular scales 
or near the plate edges, although variations in scattered light 
tend to occur over moderately large spatial scales (e.g. Figure 1).
A second, refined, reference image was constructed from the plates with 
spatially uniform zeropoints. A revised star catalogue and a 
revised set of photometric measurements were then generated. 

\subsection{Transformation into a standard system}

	The instrumental magnitudes were transformed into $V$ magnitudes, with 
the latter obtained from $G$ and $bp - rp$ entries in the GAIA DR3 using 
transformation equations from the GAIA website \footnote[2]
{$https://gea.esac.esa.int/archive/documentation/GDR2/Data_processing/chap\_cu5pho/sec\_cu5pho\_calibr/ssec\_cu5pho\_PhotTransf.html$}. 
Given the shallow nature of the two images taken with the blue emulsion, 
with the result that instrumental colors could only be found for a modest 
number of stars, an initial relation between instrumental and $V$ magnitudes 
without a color term was examined, and the result is shown in 
Figure 2. Only isolated stars that span a range of magnitudes and are 
distributed across the imaged field were considered to construct Figure 2. 
The instrumental magnitudes in Figure 2 are from the reference image described 
in Section 4.1, where sources have a higher S/N than in individual images. 

\begin{figure}
\figurenum{2}
\epsscale{1.0}
\plotone{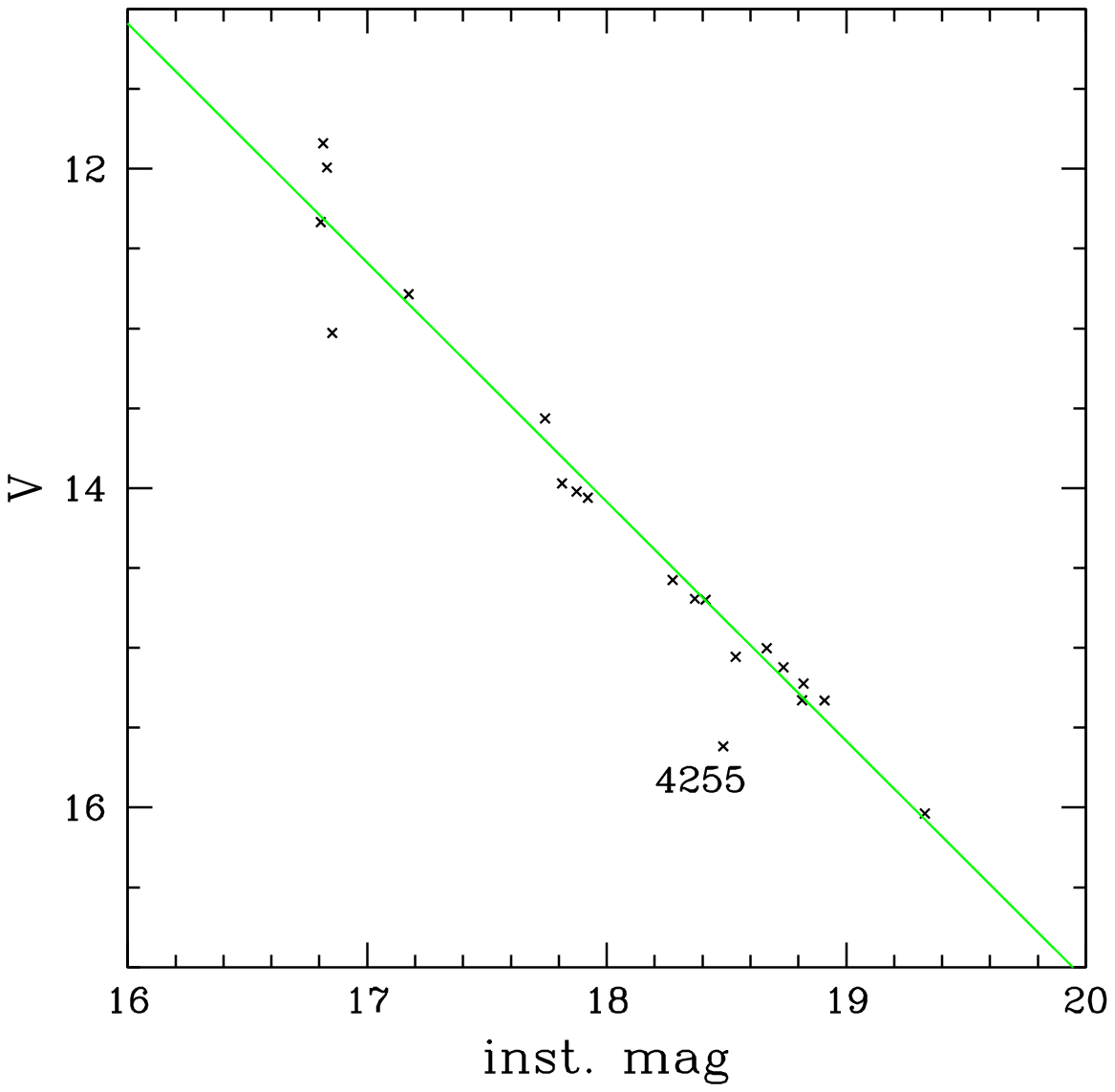}
\caption{Relation between instrumental and $V$ magnitudes. 
The green line is a linear least squares fit that does not include star 
4255, which has the smallest $bp-rp$ color of the sources in this figure. 
The scatter at the bright end is due to color 
effects, which become significant when $bp - rp \geq 1.7$.}
\end{figure}

	There is a linear relation for instrumental 
magnitudes $> 17$ in Figure 2, and the green line shows a least squares fit. 
The linearity is consistent with the majority of stars falling on the 
linear part of the characteristic curve. The stars 
with instrumental magnitudes $> 17$ in the figure 
have $bp-rp$ colors between 1.1 and 1.7. Star 4255 \footnote[3]
{The numbers assigned by DAOPHOT are used throughout this paper.}
is an obvious outlier. It has the smallest $bp-rp$ color, and was 
not included in the least squares fit.

	The linearity for instrumental magnitudes $\geq 17$ is perhaps 
surprising given the range of $bp-rp$ colors of the 
stars that fall along the relation. This linearity is likely a consequence 
of the broad wavelength response of the B emulsion, which covers the 
peak of the spectral energy distributions (SEDs) of F and G stars, coupled 
with the modest depths of absorption features in their spectra at these 
wavelengths. That there is linearity over this color range is fortunate, as 
stars within this $bp - rp$ interval account for almost 80\% of the 
sources with $G \leq 17$ in the area sampled by the plates. 
The PSF stars have instrumental magnitudes $> 17.3$, with the vast 
majority fainter than 17.5, and so fall in the linear part of this plot.

	The stars with instrumental magnitudes $< 17$ that lie above the 
green line in Figure 2 have $bp-rp$ colors in excess of 1.9, whereas 
the star that falls below the relation has $bp-rp = 1$. 
While convenient in the absence of color information, 
the relation in Figure 2 thus breaks down for stars with 
$bp - rp$ colors that fall outside the 1.1 -- 1.7 range. 
While only $\sim 2$\% of sources with $G \leq 17$ have $bp - rp > 1.8$, 
these tend to be the brightest members of M10, and many are 
either confirmed or candidate SR variables.

	Transformation relations that involve a 
wide range of $bp - rp$ colors were then examined. The behavior 
of the difference between the $V$ and instrumental magnitudes ($\Delta V$) 
$vs.$ $bp - rp$ is shown in Figure 3. Stars with $bp - rp > 2$ 
have not been included, as these tend to be variable.

\begin{figure}
\figurenum{3}
\epsscale{1.0}
\plotone{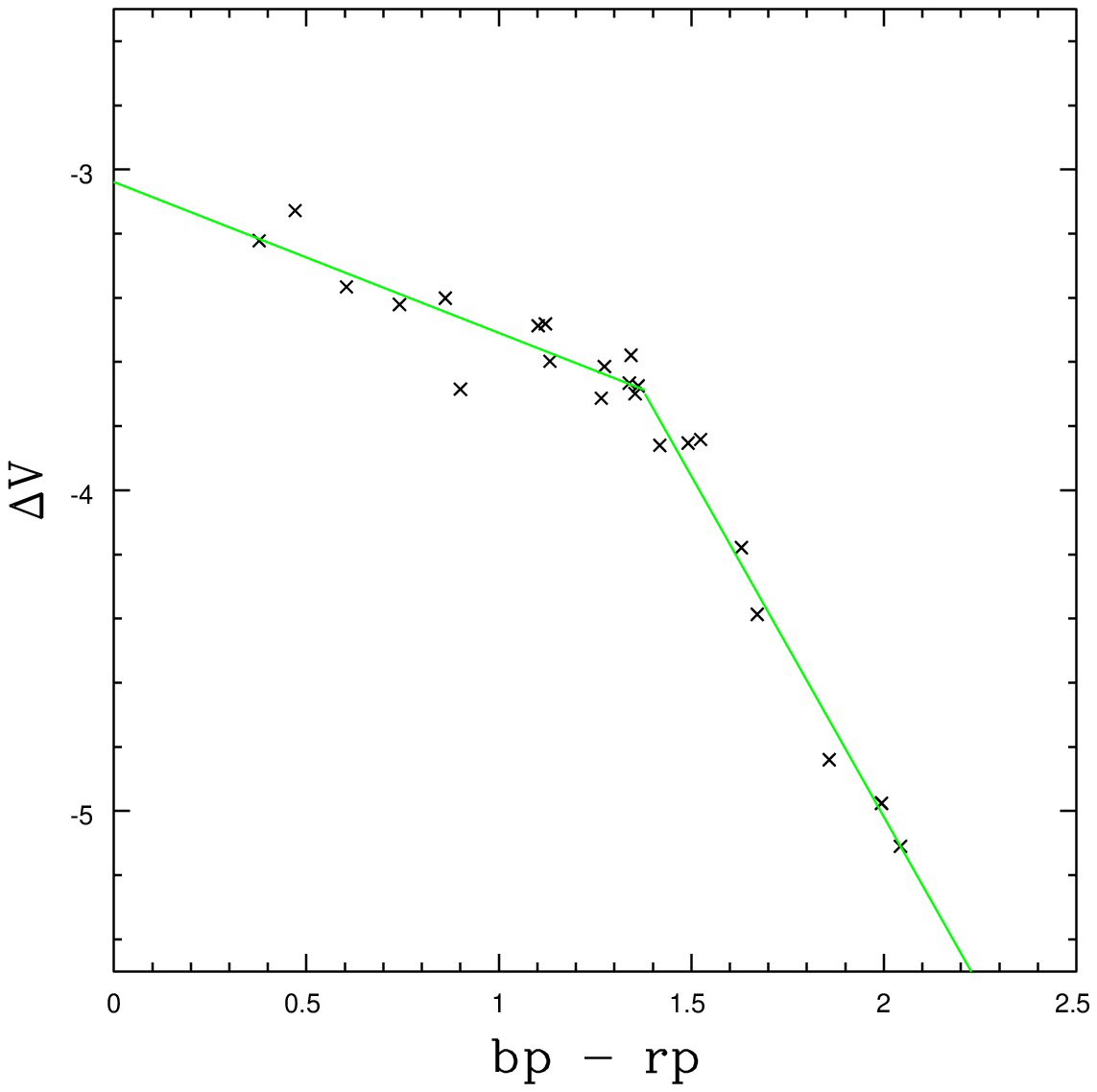}
\clearpage
\caption{Difference between $V$ and instrumental magnitudes ($\Delta V$) 
as a function of $bp - rp$. Two linear trends are defined, and the green 
lines show linear least squares fits to points with $bp-rp < 1,4$ 
and $bp-rp > 1.3$. The relations meet at $bp - rp = 1.38$.}
\end{figure}

	The points in Figure 3 define two linear relations 
with a break near $bp - rp = 1.4$. For main sequence stars this color 
corresponds roughly to late G and early K spectral types 
(i.e. the approximate point where molecular bands become important 
in spectra, and also where the peak of the SED moves out 
of the band pass of the B emulsion). The green lines in 
Figure 3 are least squares fits to points with $bp - rp < 1.4$ 
and $bp - rp > 1.3$. The overlap in $bp - rp$ between 1.3 and 1.4 
was done to better anchor an intersection 
point, which was found to occur near $bp - rp = 1.38$.

	Application of the relation in Figure 2 to variable stars in M10 
that have $bp - rp$ between 1.1 and 1.7 revealed that it reproduces well 
the expected photometric properties of stars, even those that are 
saturated. Magnitudes calculated with this relation are designated as V$_1$. 
As for stars that have $bp - rp$ colors outside of the 1.1 to 1.7 range then 
the relations in Figure 3 were applied, and those magnitudes are refered to 
as V$_2$. That this transformation procedure is applicable is demonstrated 
in Sections 6 and 7, where the transformed magnitudes are shown to reproduce 
the mean photometric properties of variable stars in M10 obtained from 
CCD measurements. 

\subsection{Plate-to-plate consistency}

	The internal consistency of the photometry 
can be gauged by examining the plate-to-plate variance in stellar magnitudes, 
and this is done in Figure 4. To be included in Figure 4 a source 
had to be recovered in all 16 B emulsion images and 
be at least 37 arcsec from the cluster center to 
avoid the areas in some plates where the signal is completely 
saturated. Some sources were found to have high dispersions 
that are skewed by a single point, and these were culled from the sample 
by applying sigma-clipping -- if a single point in the set of measurements for 
a given star had a magnitude that differed from the median magnitude of that 
source by more than $2.5\sigma$, where the variance was measured from all 16 
magnitude measurements for that source, then the source was excluded from 
Figure 4. In addition, sources at the bright end 
that are heavily saturated have large PSF-fitting uncertainties. Sources 
with uncertainties that departed markedly from the typical value 
of unsaturated sources at a given magnitude near the bright end 
were also excluded from Figure 4.

\begin{figure}
\figurenum{4}
\epsscale{1.0}
\plotone{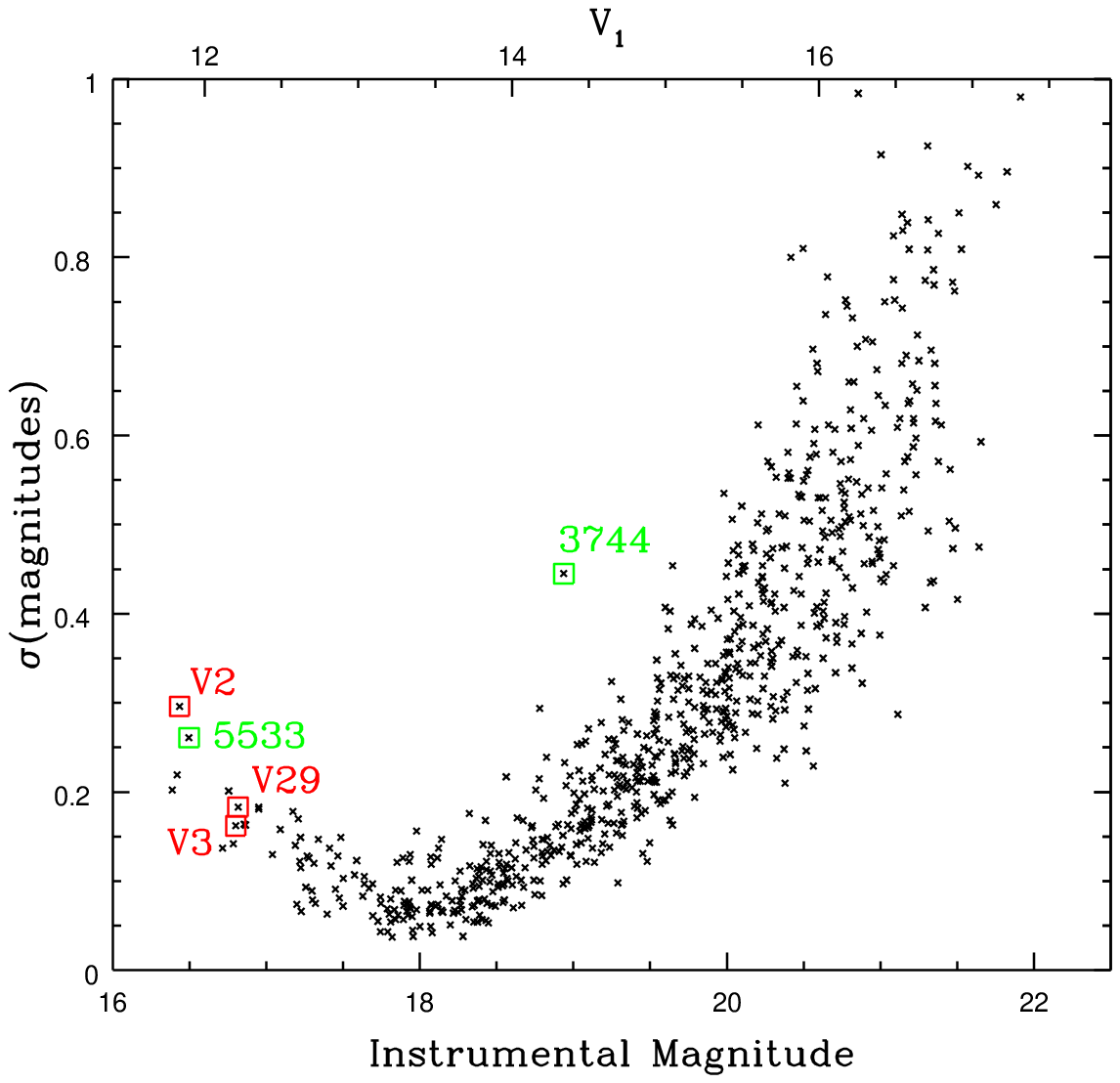}
\caption{Plate-to-plate dispersion in the photometric measurements. 
$\sigma$ is the variance in magnitudes measured from the 
16 B emulsion plates. Instrumental magnitudes are shown along 
the lower axis, while approximate $V$ magnitudes calculated from the 
relation in Figure 2 are shown along the upper axis. Sources within 
37 arcsec of the cluster center have not been included. Sources 
with a large dispersion that is driven by a single point 
or that have large uncertainties in the PSF-fitting due to saturation have 
also been excluded (see text). The upturn in $\sigma$ at the bright end is 
due to image saturation, the degree of which varies between plates. 
$\sigma$ progressively increases near the faint 
end due to the lowering of the S/N towards fainter magnitudes. 
Known variables are indicated in red, while new variables identified in 
this study are highlighted in green with their identification number. 
The W Virginis variable V2 and the new candidate variables 5533 and 3744 
stand out from the other stars on the $\sigma$ $vs.$ magnitude plane.}
\end{figure}

	The $V$ magnitudes along the top axis of 
the figure were computed using the relation in Figure 2. It should be 
emphasized that this magnitude scale does not hold for all stars in this 
figure, although it applies to the majority of sources over a broad range of 
magnitudes. The plate-to-plate dispersion for a given star does not 
depend on the transformation relation that is applied to its photometry, 
and the stars that fall well above the main concentration of points 
in Figure 4 are candidate variable stars. 
Selected known variables are indicated in Figure 4. 

	The majority of points in Figure 4 fall along a sequence with a 
width of a few hundredths of a magnitude near instrumental magnitudes 18. 
The mean dispersion along this sequence between 
instrumental magnitudes 17.5 and 18.5 is $\sim 
\pm 0.07$ magnitudes, while the dispersion is below $\pm 0.1$ magnitudes
between $V = 17$ and 19. For comparison, the scatter in 
measurements made from the DASCH archive is $\sim 0.15$ magnitude 
\footnote[1]{https://dasch.cfa.harvard.edu/dr7/introduction/}.

	The upturn at the faint end reflects the progressive lowering of the 
S/N towards fainter magnitudes, while the upturn at the bright end 
is due to image saturation. The degree of 
saturation varies from plate-to-plate, and this adds uncertainties 
to the measured magnitudes, which in turn propogate into 
the dispersion measurements. Therefore, sources that are heavily 
saturated, based on the errors computed by DAOPHOT, were excluded from 
subsequent analysis. While saturation affects some of the remaining 
objects near the bright end of the distribution in Figure 4, 
it is demonstrated in the next section that 
there is good agreement between the brightnesses 
of the variable V2 and recent CCD observations, even though V2 is at the 
bright end of Figure 4. 

	The plate-to-plate consistency in the photometry depends on the 
degree of crowding. The dispersion among magnitude measurements of 
stars within 37 arcsec of the cluster center, 
calculated from the 12 plates in which the 
central regions of the cluster are not completely saturated, 
is shown in Figure 5. The $V$ magnitude scale at the top of this figure 
is based on the relation in Figure 3 for a fixed $bp - rp =2$, 
as many of the sources near the cluster center are red giants. 
As was done with Figure 4, sources that have large dispersions 
driven by a single discrepant point and that have large PSF-fitting 
errors have been excluded from Figure 5. As expected, 
the degree of consistency at a given magnitude is poorer near the cluster 
center when compared with the measurements shown in Figure 4.

\begin{figure}
\figurenum{5}
\epsscale{1.0}
\plotone{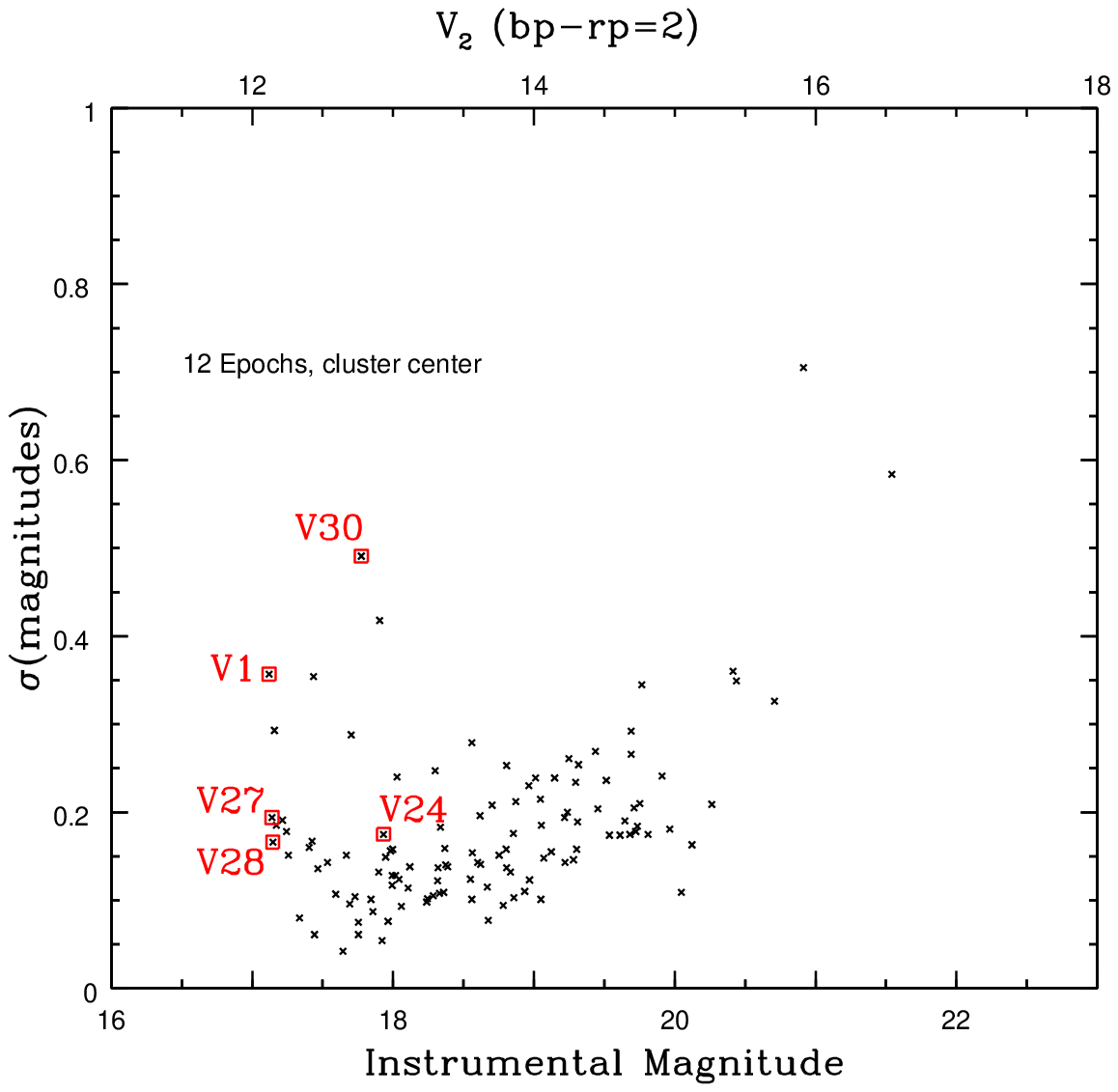}
\caption{Same as Figure 4, but for stars within 37 arcsec of the 
cluster center that are in the 12 plates where the 
area near the cluster center is not completely saturated. The upper axis shows 
approximate $V$ magnitudes calculated from the relation in Figure 
3 for a fixed $bp - rp = 2$, which is the approximate color of 
bright giants in M10. As expected, the plate-to-plate 
consistency at a given magnitude is poorer than in Figure 4 due to crowding. 
Still, internally-consistent photometry for the brightest stars at the 
10 -- 20\% level can be obtained from the plates. The variables V1 and V30
stand out with respect to the other stars.}
\end{figure}

\section{KNOWN M10 VARIABLES REVISITED}

	M10 contains a mix of variable stars. Two (V2 and V3) have 
been identified as W Vir stars, and these are among the brightest cluster 
members at visible wavelengths. W Vir stars are pulsating post-giant 
branch variables that are on the instability strip. They have 
periods that range from under a day to tens of days, and the General 
Catalogue of Variable Stars classifies stars of this type with periods 
less than 8 days as BL Her stars \citep[][]{sametal2017}. A collection of 
W Vir light curves are compiled in the OGLE \citep[][]{udaetal1993} 
Atlas of Variable Star Light Curves \footnote[1]
{https:/ogle.astrouw.edu.pl/atlas/index.html}.

	Many of the other known variable stars 
are SR variables and SX Phe stars. Both types 
of variables tend to be found near the cluster center, where 
the 4 -- 5 arcsec angular resolution of the plates presents an 
obvious obstacle for detecting all but the brightest stars. 
SX Phe stars fall on, or close to, the upper portions of the main sequence, 
and in M10 they typically have $V$ magnitudes between 17 and 18 
\citep[][]{feretal2020}, which is well below the RGB-tip. 
Given the large photometric uncertainties in objects 
with this brightness in the DAO plates (e.g. Figures 4 and 5), we will 
not consider SX Phe stars in the subsequent discussion.

	SR variables are heavily evolved post-main sequence stars, and 
are among the most luminous objects in M10. The GCVS identifies two sub-types. 
SR variables that show more-or-less regular light variations 
are of type SRa, while those that exhibit more erratic photometric 
behaviour are of type SRb. The OGLE Atlas of Variable Star Light Curves shows 
that there can be long-term trends in the mean 
light level of SR variables that are de-coupled from periodic variations.

\subsection{Variables on the Instability Strip}

	The phased light curves of the W Vir stars V2 and V3 obtained from 
the plates are shown in Figure 6. These are the most studied variables in 
M10, and so comparisons with more recent measurements are a means of assessing 
the photometry obtained from the scanned plates. Phases were calculated with 
ephemerides from Version 5.1 of the General Catalogue of 
Variable Stars \citep[][]{sametal2017}. Also shown are the 
$V$ measurements discussed by \citet{karetal2022} that 
were obtained by P. Stetson, and these will be refered to as the 'Stetson 
photometry' throughout the remainder of the paper.

	\citet{saw1938b} and \citet{arp1955} present light curves of V2 
obtained from diverse photographic sources, and those light curves are similar 
in appearance to the V2 light curve in Figure 6. There is more-or-less complete 
phase coverage of V2 in Figure 6, and the phased light curve is consistent 
with that of a pulsating variable. There is good agreement with the 
Stetson photometry, in the sense that the bright and faint 
limits of the light curve agree with the CCD measurements, 
while the scatter about the DAO light curve is also consistent with that seen 
in the Stetson photometry. This suggests that much of the scatter in the 
DAO light curve is intrinsic to the star, and is not due to uncertainties 
in the photometry. The agreement with the photometric 
boundaries defined by the CCD measurements indicates that 
saturation and non-linearity are not insurmountable obstacles to conducting 
photometry of bright stars in the DAO plates, and highlights the ability 
to extract useable light curves from them.

	There is an offset of 0.4 phase units between 
the DAO and Stetson light curves of V2 in Figure 6. 
Adopting the mid-point of the DAO observations (i.e. the summer of 1933) as 
a basis for a period calculation, then this phase 
shift is indicative of a mean period of $18.69531 \pm 
0.00003$ days over the past $\sim 80$ years, where the quoted 
error reflects a $\pm 0.02$ uncertainty in the phase difference. 
This mean period is 0.00463 days shorter than the period in the GCVS. 
The period change obtained from the digitized DAO plates is 
consistent with the period of V2 becoming shorter with time, as 
concluded by \citet{karetal2022}

	Figure 6 reveals unfortunate gaps in the 
light curve of V3, and this may explain in part why this variable 
was not detected by \citet{saw1938a}. Still, the dispersions 
in the DAO and Stetson light curves at a given phase are similar. 
Given the poor phase coverage and offset in 
mean brightnesses, no conclusions can be reached regarding a change in period 
from the DAO measurements alone. 

	With the caveat of incomplete phase coverage, 
the DAO light curve in Figure 6 is offset from 
the Stetson photometry by a few tenths of a magnitude, although the faint 
points in the DAO light curve overlap with the Stetson photometry. 
There is an offset of a few tenths of a magnitude between the mean 
$V$ magnitudes of V3 in Figures 10 and 11 of \citet{karetal2022} and Figure 3 
of \citet{feretal2020}, suggesting that the mean 
light level of V3 varies with time. Also, 
the V3 light curve obtained from blue-sensitive plates by 
\citet{arp1955} has an amplitude that is $\sim 0.3$ magnitudes larger 
than in the B CCD observations shown in Figure 10 of \citet{karetal2022}. 
Thus, there is evidence for long term variations in the photometric 
behavior of V3, as is seen in Figure 6. 

	In addition to the possible long term variations in mean 
brightness, V3 has a peculiar period for a W Vir star \citep[][]{cleetal1985}. 
\citet{sosetal2010} discuss a sample of peculiar W Vir stars, and note that 
at least some of these are in binary systems, raising the possibility that 
the properties of peculiar W Vir stars may be the result of 
binary evolution. However, there is no evidence that V3 is in a close binary 
system. The GAIA renormalized unit weight error (RUWE) is a statistic that 
is a measure of parallax reliability. This statistic is sensitive to 
binarity, as the parallax measurements change with epoch due to the 
orbital motions of the components if these motions are within the 
detection threshold of the GAIA astrometry. RUWE = 0.93 for V3 in GAIA DR3, 
indicating that if it is in a binary system then the orbital motions are 
not sufficient to affect the parallax measurements.

	The light curve of the BL Her star V24 is shown in the third 
panel of Figure 6. There are only 12 points in the DAO light curve owing to 
the saturation of the cluster core in some plates and, as was the case for 
V3, there are gaps in the phase coverage.
The $V$ light curve of V24 in Figure 3 of \citet{feretal2020} 
is well-defined with characteristics that are consistent with that of 
a pulsating star. The amplitude of the light variations in their Figure 3 
are $\sim 0.4$ magnitude in $V$, ranging between $V = 13.8$ and 14.2. 
The majority of measurements obtained from the DAO plates fall within 
these boundaries. This agreement is noteworthy given 
that V24 is located in a crowded environment, where the uncertainties 
in the photometry are larger than in the outer regions of the cluster. 

	The light curve of V22 is shown in the bottom panel of Figure 6. 
The dispersion in the DAO measurements falls within the expected $\pm 0.2$ 
magnitude range of the $V$ light curve shown in Figure 3 of \citet{feretal2020}.
When phased using the ephemeris of \citet{feretal2020} there is the hint of 
a minimum in the light curve near phase 0.3, although phase 
coverage between 0.65 and 0.85 would add confidence to the identification 
of this feature.

	\citet{feretal2020} suggest that V22 has a distance and 
kinematic properties that are consistent with M10 membership. If 
V22 is a cluster member then it would be the only known RR Lyrae 
belonging to M10. The parallax of V22 in GAIA DR3 
is $0.115 \pm 0.025$ mas. To assess distance as a criterion for membership 
in M10, the mean parallaxes of stars with $G < 13$ within 37 
arcsec of the cluster center was found; GAIA source 4365623360614730752 
was not included as it is an obvious foreground object. 
The resulting parallax for M10 is $0.198 \pm 0.010$ mas, 
indicating that V22 is in the background. It should be noted that 
the parallaxes used here are taken directly from the GAIA DR3 archive, and 
corrections for location on the sky or color have not been applied, as these 
are not expected to affect the differential comparison of parallaxes, as is 
done here. 

\subsection{SR Variables}

	Phased light curves of the SR variables V27 
and V29 are shown in Figure 7. The light curves of V1 and 
V30 are not examined because both stars are heavily blended 
in the plates. The $V$ magnitudes in this figure were 
calculated using the relation for red stars in Figure 3, with $bp-rp$ 
taken from the GAIA DR3. The ephemeris for V27 is from \citet{feretal2020}. 
These authors also estimated a period for V29, but did not cite 
a baseline epoch, and this information is also not in the GCVS. 
Therefore, the epoch used to compute phases in Figure 7 for V29 is based 
on the date of the first exposure taken in 1932.

\begin{figure}
\figurenum{6}
\epsscale{1.0}
\plotone{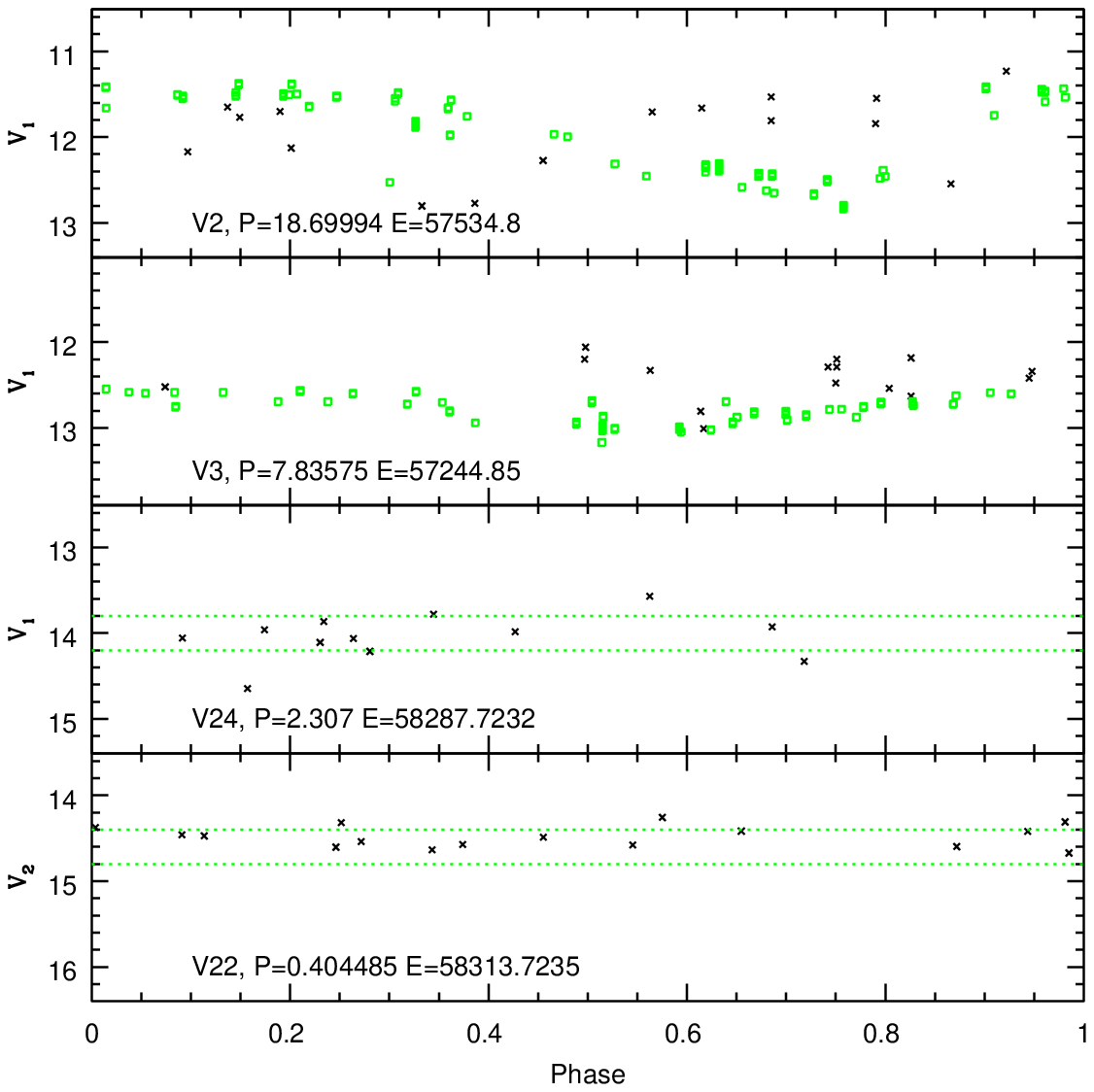}
\caption{Phased light curves of the W Vir stars V2 and V3, the BL Her star V24, 
and the RR Lyrae V22. The photometric measurements made 
from the plates are black crosses, while the Stetson $V$ 
CCD measurements are green squares. Phases were 
computed using ephemerides from the GCVS 
version 5.1 \citep[][]{sametal2017}. The relation shown in 
Figure 2 is used to compute $V$ for the three 
stars brighter than $V = 14$ (V2, V3, and V24), while the relation 
in Figure 3 was used to calculate $V$ for V22. The bright and faint 
limits of the DAO and Stetson V2 light curves are in excellent agreement. 
The $\sim 0.4$ offset in phase between the DAO and 
Stetson light curves is consistent with the period of V2 changing 
over the past century. As for V3, there is a $\sim 0.3$ 
magnitude offset between many of the DAO and Stetson 
measurements at some phases. The phase coverage of 
the DAO measurements is not sufficient to identify a difference in phasing. 
Still, the dispersion in the DAO measurements at a given phase is comparable to 
that in the Stetson photometry. The phase coverage of the V24 measurements 
is poor, although the bright and faint limits of 
the light curve in Figure 3 of \citet{feretal2020}, shown with the dashed green 
lines, are consistent with the majority of the DAO measurements. 
V22 is a RR Lyrae variable that has been associated with M10 
\citep[][]{feretal2020}, and its light variations are consistent with 
the $\pm 0.2$ magnitude variations seen in the DAO light curve. There is 
also good agreement with the bright and faint limits in Figure 3 of 
\citet{feretal2020}.}
\end{figure}

\begin{figure}
\figurenum{7}
\epsscale{1.0}
\plotone{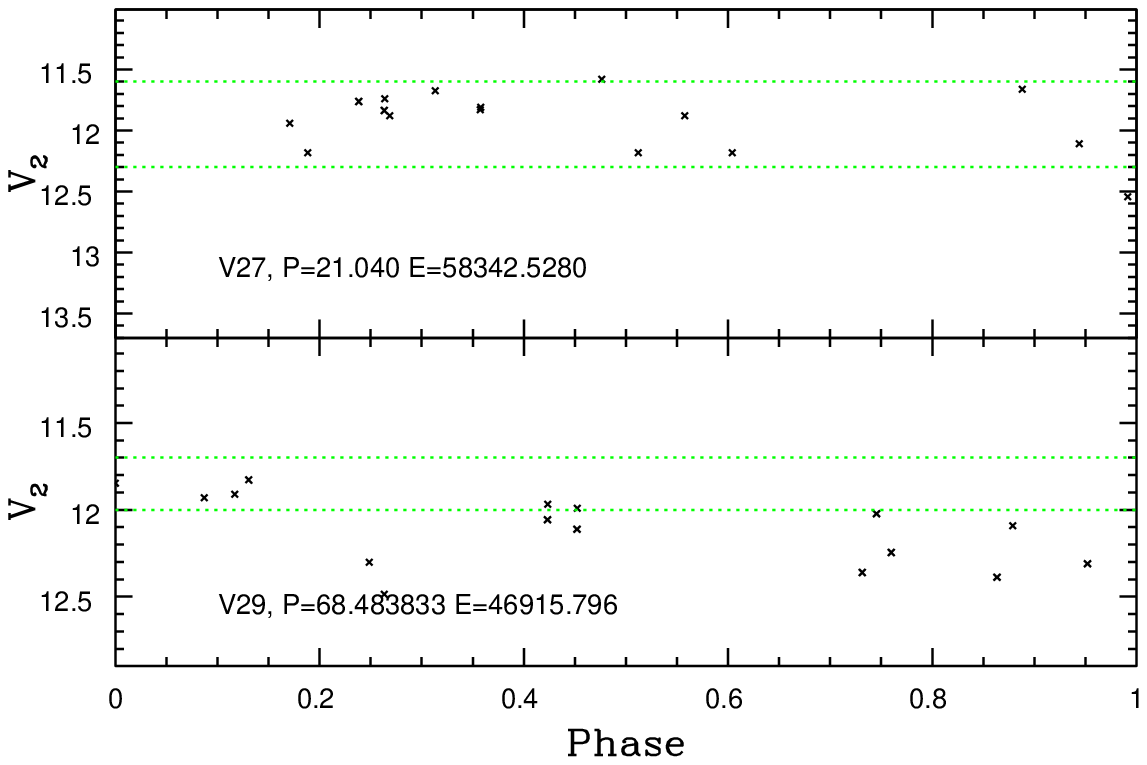}
\caption{Phased light curves of the SR variables V27 and V29. 
The green dashed lines are the approximate bright and faint limits of 
the light curves in Figure 3 of \citet{feretal2020}. 
The periods used to phase the data are from \citet{feretal2020}. 
We are not aware of a published epoch for V29, and the phasing used 
here assumes a baseline epoch of July 26, 1932, which is the date of the 
first observation taken with the 'B' emulsion. The DAO measurements for V27 
fall within the bright and faint limits defined 
by CCD light curves. As for V29, some of the DAO measurements fall within 
the limits defined by the CCD data, although there is a systematic downward 
trend with increasing phase, suggesting that the measurements 
discussed by \citet{feretal2020} may not track the entire light cycle.} 
\end{figure}

	The location of V27 on the \citet{feretal2020} CMD of M10 places 
it near the RGB-tip, and this is consistent with its red color in the GAIA 
DR3 archive ($bp - rp = 2.14$). The majority of the DAO photometry 
of this star falls within the bright and faint limits of the CCD light curve 
presented by \citet{feretal2020}, and so there is no evidence for a long term 
trend in the mean $V$ magnitude. With the caveat that the phase coverage of the 
DAO light curve is not complete, there is no evidence for 
the bow-shaped behavior that is apparent in the \citet{feretal2020}
light curve.

	The \citet{feretal2020} CMD of M10 indicates that V29 is close 
to the RGB-tip. The parallax, proper motions, and velocity of V29 taken 
from GAIA DR3 are consistent with membership in M10. However, it 
is offset by 6.5 arcmin from the cluster center, which corresponds to 
$\sim 9$ parsecs at the distance of M10, placing it well outside of the 
half light radius. 

	The \citet{feretal2020} light curve of V29 consists of three 
data clumps, with a $\sim 0.2$ magnitude variation between these, 
and an internal scatter of $\pm 0.05$ mag within each clump. 
While some of the points in the DAO light curve of V29 fall within 
the maximum and minimum limits in the \citet{feretal2020} 
light curve, there is a systematic trend in the DAO measurements 
indicating a minimum near phase 0.8 that is a few tenths of a 
magnitude fainter than the \citet{feretal2020} minimum brightness. Given the 
spotty phase coverage for V29 in Figure 3 of \citet{feretal2020} 
then this suggests that their photometry does not 
sample the full photometric behavior of this star.

\section{OTHER VARIABLES}

\subsection{New Variables Found From the Plates}

	There have been recent surveys for 
variable stars in and around M10, and so it is likely that any 
remaining moderately bright variables that have yet to be discovered will 
be difficult to detect due to low amplitude variations and/or very 
long periods, given that the most obvious have already been found. 
The crowded inner regions of M10 are also difficult to explore 
with ground-based images that sample the local seeing, and the use of 
photometric measurements of individual stars as a variability diagnostic 
near the cluster center will only trigger on the brightest stars. 
While there are undoubtedly variable stars that await 
detection near the center of M10, we focus here on the area outside 
of the crowded central regions.

	There are a number of outliers in 
the dispersion $vs.$ magnitude diagrams in Figures 4 and 5, and 
some of these may be new variables. Variance is an admittedly 
blunt tool for identifying variables, as stars with one or two measurements 
that depart from the other observations of a star due to -- say -- cosmetic 
flaws in the plates will have a large dispersion. 
However, such points have been suppressed in Figures 4 and 5.
A more problematic issue is that {\it bona fide} 
variables with modest light variations will go undetected.

	The search for new candidate variables was restricted to $V < 
16$, as measured in the median image. At fainter magnitudes 
the plate-to-plate dispersion balloons, making the detection of variables 
problematic. While there are outlier points in Figure 5, 
it was decided not to consider these as new variables given the modest number 
of points, and the larger inherent scatter in the photometry near the cluster 
center.

	Two new candidate variables, stars 3744 and 5533, were identified 
based on their locations in Figure 4. The former is a conspicuous outlier 
from other stars with the same brightness, while the latter has a location 
on the $\sigma$ $vs.$ magnitude plane that is similar to that of V2, 
for which a light curve that is consistent with more modern measurements 
was constructed (Figure 6). The location, photometric properties, and parallax 
of each star are listed in the top rows of Table 3. Neither star is 
in the GCVS. However, after star 3744 was flagged as a variable 
based on its location in Figure 4 we became aware 
that it had been been identified independently 
as a variable in the GAIA DR3 archive. 

	With the exception of the last column, the entries in Table 3, 
including the right ascension and declination, are those in the GAIA DR3 
archive \footnote[1]{https:gea.esac.esa.int/archive/. The procedures for 
measuring the entries in Table 3 have been described by \citet{gai2023}, 
\citet{katetal2023}, and \citet{linetal2021}.} The 
parallaxes of star 3744 and M10 agree at roughly the 
$1\sigma$ level. The $bp - rp$ color of star 3744 is far larger than that 
of stars near the M10 RGB-tip, which typically have $bp - rp \sim 2$.
Star 3744 is 2MASS source 16574980--0403150, with $K=6.8$ and 
$J-K = 1.293$ in the 2MASS Point Source Catalogue \citep[][]{skretal2006}.
If it is a member of M10 then it is likely the most luminous object in 
the cluster. 

	The location of star 3744 on the CMD of M10 provides possible 
insights into cluster membership. The M10 CMD constructed from entries in 
the GAIA database is shown in Figure 8. To suppress 
contamination from field stars, only photometric measurements of stars 
that fall within 90 arcsec of the cluster center, which is the 
approximate half light radius, are shown in the figure. 
Additional constraints were also placed on the 
parallax and proper motions, and these are specified in the figure.

	The location of Star 3744 on the CMD 
is offset from the cluster sequence along the color axis by more 
than two magnitudes, while the Gp brightness is over 1 magnitude 
below the giant branch tip as defined by the brightest giants 
in the main body of the cluster. Given that star 3744 is likely highly evolved 
then its photometric properties at visible and red wavelengths are 
likely subject to significant circumstellar extinction as well as 
line blanketing. Still, as star 3744 has photometric properties that are 
very different from any other star in M10 and is located well outside 
of the main body of the cluster then there is a good chance that it is 
a field star.

\begin{figure}
\figurenum{8}
\epsscale{1.0}
\plotone{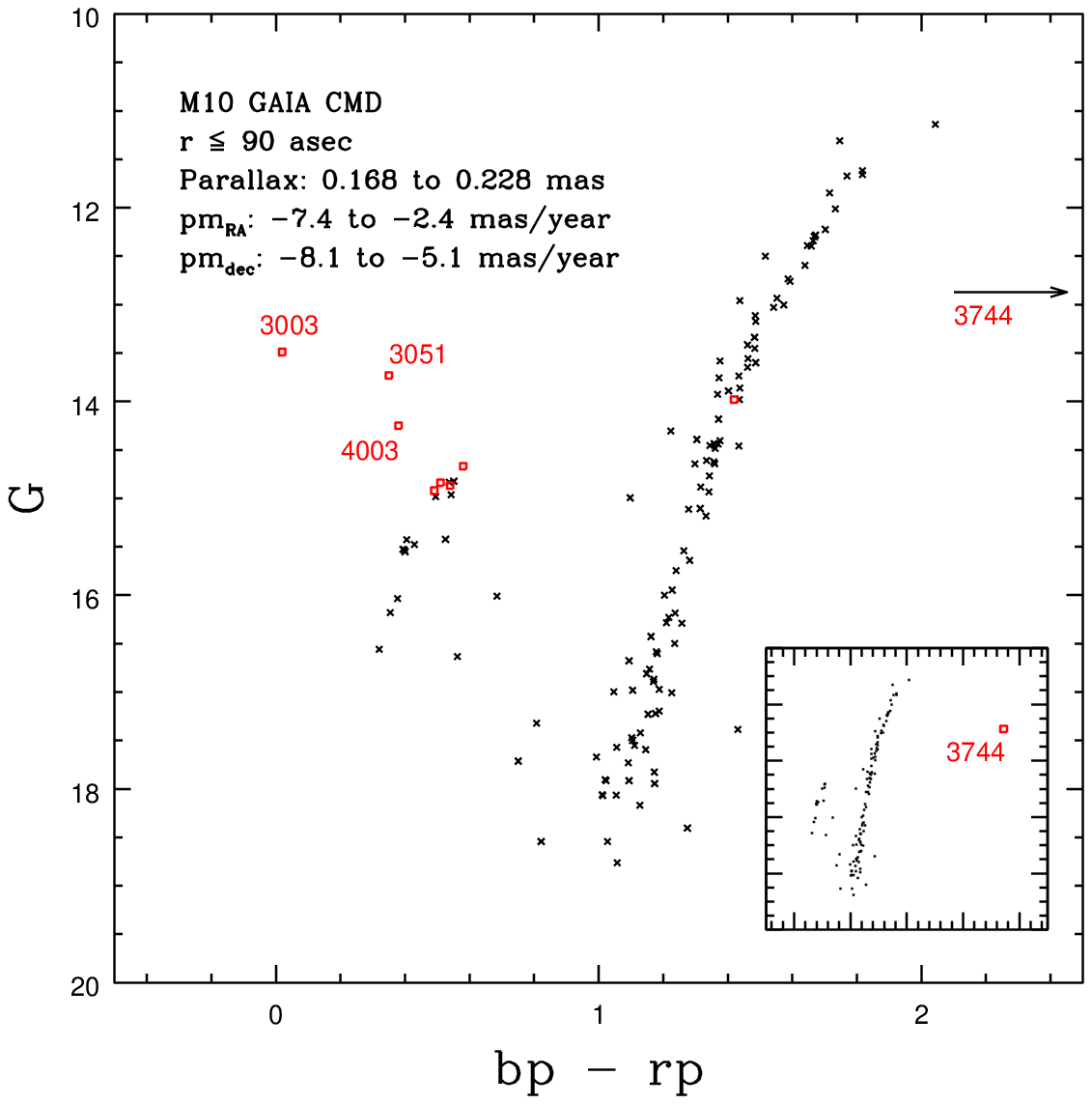}
\caption{CMD of stars in M10 that are within 90 arcsec of the cluster 
center, which corresponds roughly to the cluster half light radius.
The photometric measurements are from GAIA DR3. 
Additional constraints were placed on the parallax and proper 
motion measurements to suppress contamination from non-cluster stars. 
The open red squares are stars in Tables 3 and 4 that have parallaxes 
$< 0.5$ mas. Stars that are outliers from the cluster sequence are labelled.
Star 3744, which was identified as a variable star, falls well off of the M10 
giant branch, and has a Gp that is over 1 magnitude below the giant branch tip 
as defined by the brightest giants near the cluster center. The location 
of star 3744 with respect to the M10 CMD is indicated in the inset.} 
\end{figure}

	The light variations of star 3744 are examined in Figure 9. 
The $V$ magnitudes of this star were estimated using the 
relation in Figure 3, although we caution that this relation is 
calibrated only up to $bp - rp = 2$. Hence, the 
$V$ magnitudes should be considered as provisional only.

\begin{figure}
\figurenum{9}
\epsscale{1.0}
\plotone{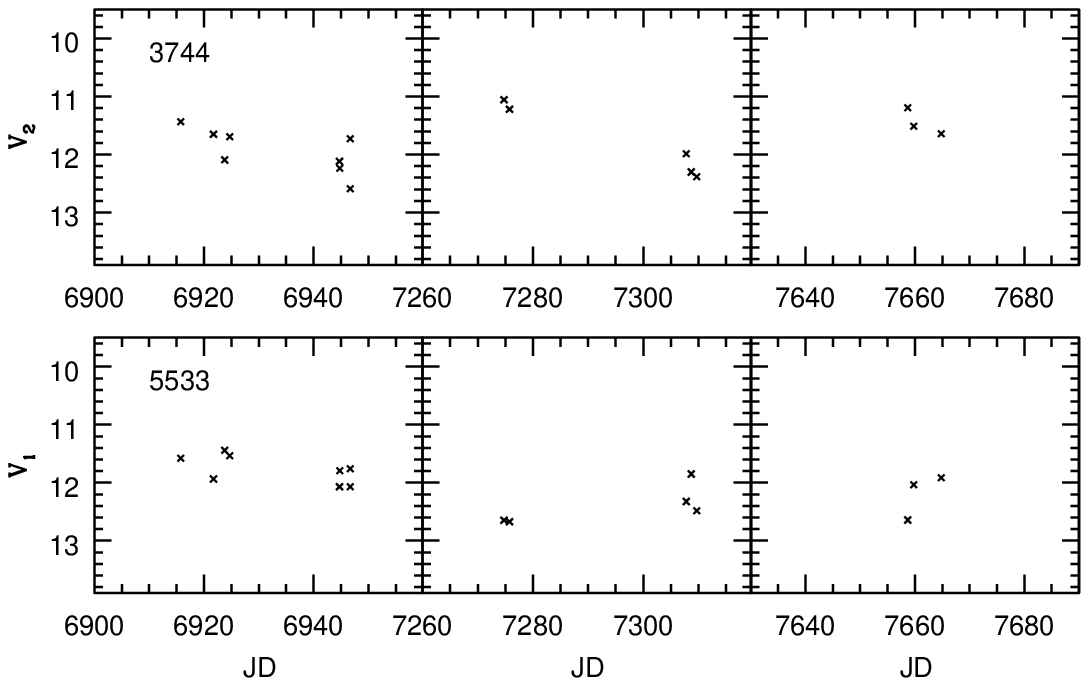}
\caption{Light variations of the stars 3744 and 5533, which were identified as 
variable based on their locations in Figure 4. The parallax of star 3744 
differs from that of M10 at roughly the $1\sigma$ level. Its photometric 
properties are consistent with it being a highly evolved star, and its very 
red color places significant uncertainties on the estimated $V$ magnitudes. 
While time coverage is spotty, the repeatability of the points 
is suggestive of a period near $\sim 80$ days. Star 3744 then appears to be 
an LPV at roughly the same distance as M10. The parallax of star 5533 
places it in the foreground, and its photometric properties suggest that 
it is likely a K giant.} 
\end{figure}

	While the time coverage of 3744 is spotty, the light variations are 
consistent with that of a pulsating variable with an apparent period of 
$\sim 80$ days and an amplitude of $\sim 1$ magnitude. 
When taken together, the photometric properties 
of star 3744 are consistent with it being a long period variable (LPV). 
We suspect that it is either (1) an LPV in the field 
that may be physically close to M10, or (2) an 
LPV that is an outlying member of M10. LPVs have been discovered in 
clusters that are more metal-poor than M10 \citep[][]{sahetal2014}, 
and so either of these is possible.

	The parallax of star 5533 places it in 
the foreground, while its proper motions are also 
markedly different from those of cluster members. Based on its 
color, brightness, and parallax, we suspect that it is a K giant. The 
light curve in Figure 9 shows photometric variations on the order of 
1 magnitude. Photometric variations among single K giants 
are rare \citep[e.g.][]{per1993}, and the variations seen here may be 
indicative of an eclipsing system. While there is no clear sign of 
periodicity, this may change with additional photometric measurements. 

\subsection{Assessing Variables Flagged in the GAIA Database}

	\citet{eyeetal2023} discuss the properties of variable stars 
in GAIA DR3. We have examined the light variations of the sources flagged 
as variable in GAIA DR3 that fall within the area covered by the DAO plates 
and that have $G$ between 13 and 15. Fainter stars 
were not examined as the plate-to-plate dispersion increases to a point that 
only loose limits can be placed on the amplitude of photometric variations.

	GAIA variables were identified on 2MASS and Digitized Sky Survey 
images using the right ascensions and declinations from the GAIA DR3. 
While these datasets were recorded at epochs 
that are intermediate between when the DAO plates 
and the GAIA observations were obtained, in all cases a source with 
the approximate expected brightness of the variable was found near the 
expected location. This is not unexpected as the proper motions of the 
variables are only a few mas year$^{-1}$, and so the overall motion on the 
sky over the eight decades between when the DAO plates were the GAIA data were 
recorded is less than an arcsec.

	The properties of the GAIA variable stars in the DAO plates 
are summarized in Tables 3 ($bp - rp > 1$) and 4 $(bp - rp < 1)$. 
The stars in these tables have been grouped according to color given 
potential differences in the nature of the 
objects. Some of the bluer variables may be RR 
Lyraes, as they have $G$ and $bp - rp$ similar to that of V22. 
The stars in each table are ordered according to projected distance 
from the cluster center.

	None of the stars in Tables 3 and 4 are in the GCVS, nor are they 
listed in Table 3 of \citet{feretal2020}. With the exception of the 
last column of these tables, which gives the plate-to-plate dispersion in 
the photometry, the information in these tables comes from the GAIA DR3. 
Three of the stars in Table 4 (\#s 4836, 4003, and 5314) have parallaxes 
that agree with that of M10 at the $1\sigma$ level.
The majority of the sources have plate-to-plate 
dispersions that are consistent with intrinsic variations at no more than 
the 10\% level.

\begin{deluxetable}{lcccccccc}
\tablecaption{Observational Properties of Candidate Variables With 
$bp - rp > 1$}
\tablehead{ID\tablenotemark{a} & RA & Dec & $G$ & $bp-rp$ & $\pi$ & pm$_{RA}$ & pm$_{Dec}$ & $\sigma$\tablenotemark{c} \\
 & (J2000) & (J2000) & & & (mas) & (mas/year) & (mas/year) & (mag) \\}
\startdata
46553439247616(3744) & 16:57:49.8 & --04:03:15.0 & 12.87 & 3.72 & $0.159 \pm 0.045$ & --5.82 & --4.88 & 0.445 \\
22192383394304(5533) & 16:57:18.7 & --04:09:06.4 & 11.15 & 1.43 & $0.959 \pm 0.026$ & 0.86 & --6.64 & 0.261 \\
 & & & & & \\
23291895153920(4367)\tablenotemark{b} & 16:57:09.4 & --04:07:14.1 & 13.98 & 1.42 & $0.147 \pm 0.015$ & --4.90 & --6.53 & 0.101 \\
19443604279168(6071) & 16:57:41.9 & --04:08:52.4 & 13.18 & 1.05 & $2.117 \pm 0.016$ & --10.59 & --7.02 & 0.170 \\
34493170387840(2262) & 16:56:29.4 & --04:05:21.4 & 14.18 & 1.13 & $1.913 \pm 0.023$ & --6.12 & --5.16 & 0.127 \\
\enddata
\tablenotetext{a}{GAIA source ID - 4365600000000000000. The ID assigned by DAOPHOT is in brackets.}
\tablenotetext{b}{12 epochs only.}
\tablenotetext{c}{Disperson in photometric measurements.}
\end{deluxetable}

\begin{deluxetable}{lcccccccc}
\tablecaption{Observational Properties of Candidate Variables With 
$bp - rp < 1$}
\tablehead{ID\tablenotemark{a} & RA & Dec & $G$ & $bp-rp$ & $\pi$ & pm$_{RA}$ & pm$_{Dec}$ & $\sigma$\tablenotemark{c} \\
 & (J2000) & (J2000) & (mag) & (mag) & (mas) & (mas/year) & (mas/year) & (mag) \\}
\startdata
35244782836096(3051)\tablenotemark{b} & 16:57:09.8 & --04:04:28.6 & 13.73 & 0.35 & $0.166 \pm 0.021$ & --4.34 & --6.25 & 0.151 \\
35249084745600(3003)\tablenotemark{b} & 16:57:09.2 & --04:04:24.5 & 13.49 & 0.02 & $0.113 \pm 0.023$ & --4.84 & --6.79 & 0.098 \\
21573908200448(4836) & 16:57:07.7 & --04:08:17.0 & 14.87 & 0.54 & $0.200 \pm 0.028$ & --5.21 & --6.53 & 0.068 \\
33496737712768(4003) & 16:56:57.6 & --04:07:16.1 & 14.25 & 0.38 & $0.218 \pm 0.023$ & --4.79 & --6.74 & 0.128 \\
21672686027264(4705) & 16:57:00.1 & --04:08:34.7 & 14.84 & 0.51 & $0.121 \pm 0.029$ & --4.41 & --6.25 & 0.060 \\
21054208241408(5314) & 16:56:59.5 & --04:09:53.7 & 14.92 & 0.49 & $0.185 \pm 0.030$ & --4.72 & --6.65 & 0.038 \\
36108077928576(1710) & 16:57:16.5 & --04:00:45.7 & 14.67 & 0.58 & $0.128 \pm 0.031$ & --4.81 & --6.67 & 0.037 \\
34768048188416(2124) & 16:56:46.0 & --04:03:53.4 & 13.78 & 0.73 & $0.836 \pm 0.030$ & --2.14 & 0.47 & 0.089 \\
\enddata
\tablenotetext{a}{GAIA source ID - 4365600000000000000. The ID assigned by DAOPHOT is in brackets.}
\tablenotetext{b}{12 epochs only.}
\tablenotetext{c}{Disperson in photometric measurements.}
\end{deluxetable}

\subsubsection{Assessing Cluster Membership}

	\citet{feretal2020} discuss the proper motions of 
stars near M10 using information from GAIA DR2, and the distribution of objects 
in the pm$_{RA}$ $vs.$ pm$_{dec}$ plane is shown in their Figure 2.  
Stars in M10 are concentrated around pm$_{RA} = -4.9$ 
mas yr$^{-1}$ and pm$_{dec} = -6.6$ mas yr$^{-1}$, 
forming a distribution that extends over $\sim 10$ mas yr$^{-1}$ 
in pm$_{RA}$ and $\sim 6$ mas yr$^{-1}$ in pm$_{dec}$. 
Field stars make up a diffuse cloud of points in the proper 
motion plane that, while offset slightly from stars in M10, 
overlaps with much of the cluster distribution. The proper motions 
of almost all of the stars in Tables 3 and 4 place them in 
the area occupied by cluster members on the proper motion plane. The 
exceptions are the three stars with parallaxes indicating that they are 
in the foreground.

	Radial velocities and location on the CMD provide additional 
information about cluster membership. Only 2 of the stars in Table 3 and 
none of the stars in Table 4 have velocities in GAIA DR3. The two 
stars that have velocities are clearly foreground objects.
As for location on the CMD, there are obvious outliers that are labelled in 
Figure 8.

	Stars in Tables 3 and 4 that have parallaxes 
$< 0.5$ mas are marked with red squares in Figure 8. Stars 4836, 4705, 
5314, and 1710 fall squarely on the upper regions of the cluster HB. 
While star 3003 has a parallax that differs from that of M10, 
this is not the case for stars 3051 and 4003. Star 3051 is 
located near the cluster center and it, along with star 
4003, falls in a part of the CMD that is occupied by supra-horizontal 
branch stars in globular clusters that have [Fe/H] 
similar to that of M10 \citep[e.g.][]{shaetal1998}.

\subsubsection{Light Curves}

	Figure 10 shows the location of the GAIA variables on the 
$\sigma$ $vs.$ magnitude diagram. The upper panel of Figure 10 is for sources 
that are more than 37 arcsec from the cluster center, while the lower panel is 
for sources that are within this area. Many of the GAIA variables fall along 
the lower envelope of the distributions in Figure 10, indicating that the 
amplitude of light variations is at or below the 10\% level. 
Such a modest amplitude is likely a selection effect, as variables 
with larger amplitudes may have preferentially been found 
in previous surveys. Still, there are three GAIA 
variables that fall on or near the upper envelope of the locus in the upper 
panel of Figure 10. These are stars 6071, 4003, and 2262, and their 
light curves are shown in Figure 11.

\begin{figure}
\figurenum{10}
\epsscale{1.0}
\plotone{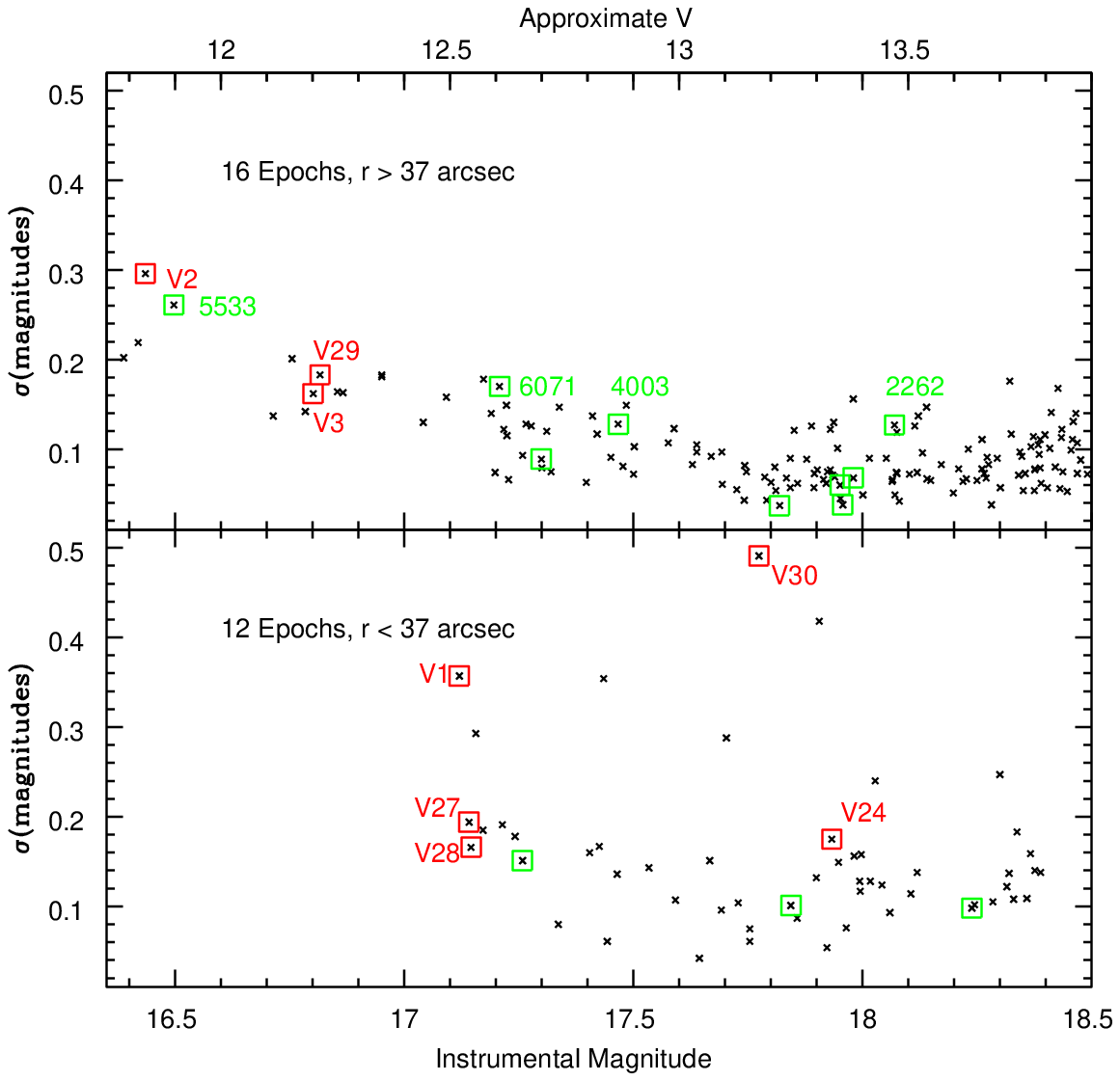}
\caption{Location on the $\sigma$ $vs.$ magnitude plane of variables in the GAIA 
DR3. The upper panel includes sources that are at least 37 arcsec from the 
cluster center, while the lower panel shows sources within 37 arcsec 
of the center. The GAIA variables and new candidate 
variables are marked with green squares. Only 
three of the stars flagged as variable in the GAIA database (6071, 
4003. and 2262) fall along or close to the upper envelope along the 
dispersion axis, and so are candidates for finding photometric variations. 
Star 4003 has a parallax that is consistent with that of M10.}
\end{figure}

\begin{figure}
\figurenum{11}
\epsscale{1.0}
\plotone{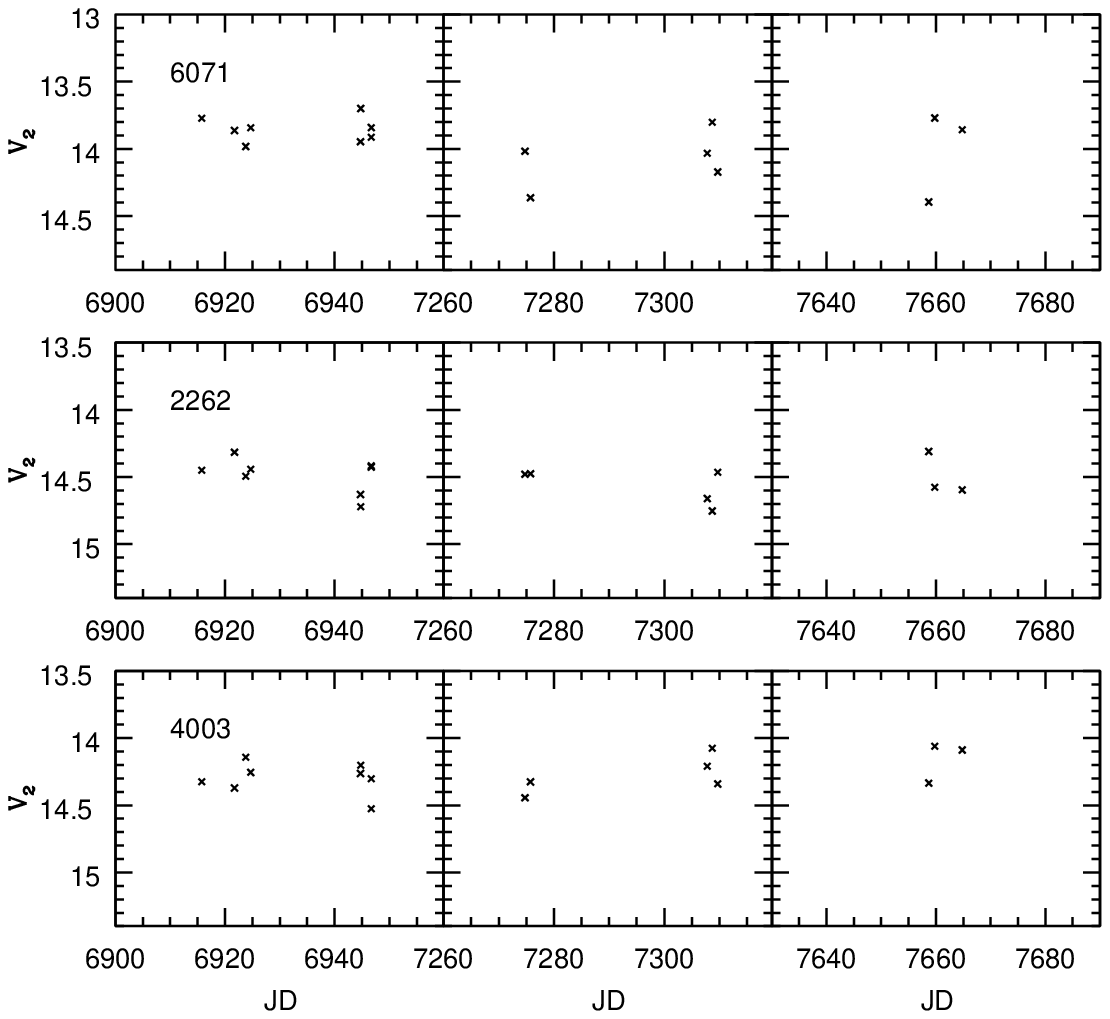}
\caption{Photometric observations of the stars 6071, 2262 and 4003, which 
are listed as variable in the GAIA DR3. Evidence for variablity 
with an amplitude of roughly 1 magnitude can be seen in the light 
curve of the foreground star 6071. There is 
evidence for long-term and possible short-term variations. Like star 6071, 
star 2262 has a parallax that is consistent with it being a foreground object. 
The dispersion during the same observing blocks suggests that 
variations may occur on time scales of a day or less.
Star 4003 has a parallax that is consistent with that of M10. 
It is offset from the cluster HB on the CMD by roughly 
1 magnitude, suggesting that it might be a supra-HB 
star. The dispersion may be due to long-term variations, 
although it falls close to the ridgeline along the $\sigma$ axis in Figure 10.}
\end{figure}

	Star 6071 has the second highest $\sigma$ among the GAIA 
candidate variables, and its parallax indicates that it is a foreground 
object at a distance of $\sim 0.5$ kpc. The 
$bp - rp$ color is consistent with a G spectral type. 
Evidence for variablity is clearly evident in the light curve, where 
the amplitude of the variations is roughly 1 magnitude. There is 
evidence for long-term, and possibly short-term, variations.

	Like star 6071, star 2262 has a parallax 
that is consistent with it being a foreground object. 
The dispersion in brightness during the same observing blocks suggests that 
photometric variations may occur over time scales of a day or less. Based 
on the limited number of observations in Figure 11, no conclusion can 
be reached as to the type of variability.

	Star 4003 has a parallax and proper motions that 
are consistent with that of M10. It falls close to the ridgeline 
along the $\sigma$ axis in the upper panel of Figure 10, suggesting that 
the amplitude of light variations is likely modest. 
There are hints of possible long-term variations in 
the light curve, although the nature of any variations are 
unclear from these data. The $G$ and $bp - rp$ of 
star 4003 are similar to those of V22, raising the possibility that it 
might be an RR Lyrae variable. However, it falls almost 1 magnitude 
above the cluster HB in Figure 8, suggesting that if it is a cluster 
member then it could be evolving on the supra-horizontal branch. 
Assuming that it is equidistant with M10 then it 
has a projected offset of 4.5 pc, or just over twice the 
cluster half light radius, from the cluster center. 

\section{DISCUSSION \& SUMMARY}

	We have used an Epson 12000XL scanner to digitize plates of 
the globular cluster M10 that were recorded with the DAO 1.8 meter telescope 
during the 1930s. Other groups have used the same, or 
similar, model of scanner to digitize plates in their collections 
\citep[e.g.][]{ceretal2021, aoketal2021}. When combined with more recent 
observations, the DAO plates can extend the 
time coverage of known variables in and around 
M10 by almost a century into the past. Lessons learned from 
this study should be applicable to similar plate collections.

	Measuring stellar brightnesses from photographic plates is 
hardly new. However, the significance of the present work is that many of the 
DAO plates are not of high quality, but are likely more-or-less typical of 
what would have been recorded for targeted scientific use during the 1930s, as 
opposed to a dedicated survey where homogeneity should be a greater concern. 
There are obvious issues with scattered light, grime, cosmetic flaws, 
writing, and ink blots. The plates were also recorded at moderately high 
airmass, and so variations in sky transparency and seeing are potential 
concerns. When considered in concert, these factors have the potential to 
compromise the plate-to-plate photometric consistency that is needed to 
search for and examine all but the brightest variable stars. 

	\citet{karetal2022} conducted a study of 
three bright M10 variables using information from 
these and other plates, but considered only published times of 
maximum/minimum light; the plates were not digitized to extract light curves. 
The digitization of plates has the potential to produce refined photometric 
measurements that will allow more reliable information for assessing 
changes in the characteristics of variable stars, 
while also allowing fainter variables than those identified from visual 
inspection alone, to be examined. The ability to work in areas of high stellar 
density, where blending can confound by-eye measurements, is also enabled.

	The non-linear response of photographic plates 
compromises the ability to define a PSF that holds over a 
wide range of magnitudes \citep[e.g.][]{ste1979}. Nevertheless, after 
balancing the photometric response across the plates and performing 
PSF-fitting in which the fit is extended well into 
the PSF wings, photometry is obtained that is 
{\it internally} consistent (i.e. from plate-to-plate) to better than 10\%. 
This falls short of the state-of-the-art 
photometric measurements that were obtained towards the 
end of the wide-spread use of photographic plates in the 1970s - 1980s
\citep[e.g.][]{hesandhar1977, chuandfre1978}. However, 
high quality measurements of this nature were obtained from plates 
that were recorded during good observing conditions, 
with the specific intent of obtaining photometry at the few percent level. 
The emulsions available at that time were also the product 
of refinement over the many decades since the DAO plates were recorded. 

	The transformation into a standard system 
for comparison with more recent observations is a 
source of uncertainty. The wavelength coverage of the 
$V$ filter approximates that of the 'B' emulsion, and 
two transformation relations between instrumental and $V$ 
magnitudes have been explored. Evidence is presented that 
stars with saturated light profiles and $bp-rp$ between 1.1 and 1.7 can be 
transformed into $V$ with uncertainties of only a few percent without 
resorting to a color term. That such a relation holds is likely due to the 
overlap of the $V$ filter and B emulsion wavelength coverages, coupled with the 
shape of the SEDs of stars in this color range within the filter bandpass.
A transformation relation that involves a color term for other stars was also 
found. This relation relies on $bp-rp$ colors, as the two blue plates 
of M10 do not go deep.

	Light curves of known variable stars have been constructed, 
and these include the W Vir stars V2 and V3. \citet{karetal2022} 
examined the periods of both stars over a long time baseline, and found 
that the periods of both are increasing with time. 
The DAO light curve of V2 is consistent with this 
conclusion, and we find a mean period of $18.69531 \pm 0.00003$ days 
over the past eight decades.

	As for V3, \citet{karetal2022} used phase information from 
\cite{cleetal1985}, relying mainly on published points of maximum light 
to examine period changes. Figure 14 of \citet{karetal2022} shows 
substantial scatter in the phase shift $vs$ time diagram of V3. Information 
from the DAO plates for V3 was not considered by \cite{karetal2022} 
as \cite{saw1938a} did not detect this variable. This is unfortunate, as 
the DAO light curve of V3 hints at long term photometric trends 
for that star, in the sense that the DAO measurements are offset in brightness 
from what might be expected from more recent observations. The 
mis-match in the light curve that we construct for V3 with respect 
to more recent observations demonstrates that it would be worthwhile 
to re-visit the behaviour of this and other variables 
that have modest light variations by examining digitized plates.
It is also noted that V3 may be a peculiar W Vir star, 
of the type discussed by \citet{sosetal2010}. A search for a close binary 
companion would be useful to further explore this possibility. 

	The photometry of two other stars on 
the instability strip (the RR Lyrae star V22 and the BL Her star 
V24) has also been discussed. While \citet{feretal2020} argue that V22 might 
be the only RR Lyrae star in M10, the parallax in the GAIA DR3 suggests that 
it may be a background object. The scatter in the photographic photometry and 
the gaps in the phase coverage of both variables complicate efforts 
to examine long-term trends in periodicity. However, the 
dispersion in the measurements obtained for these variables 
from the DAO plates is consistent with that seen in 
recent measurements. We find no evidence for a systematic change 
in the mean brightness of these stars.

	Photometric measurements for selected known SR variables 
have also been recovered. The light estimates 
obtained for these variables from digitized plates are of interest as 
these stars have long periods that complicate 
efforts to probe long-term photometric trends.
The phase coverage then tends to be spotty \citep[e.g.][]{feretal2020}. 
The photometric measurements of SRs obtained from the DAO plates 
have dispersions and mean brightnesses that are consistent with 
those made with CCDs. In the case of V29 the phase coverage is sufficient 
to conclude that the photometric variations may have a larger amplitude 
than those estimated from more recent studies.

	The low quantum efficiency of photographic plates 
limits their ability to go deep, even when compared 
with exposures taken with small telescopes that are 
equipped with CCDs. Despite this limitation, two candidate variables that are 
not listed in the current edition of the GCVS have been identified. 
One star, 3744, was independently designated as a 
variable in the GAIA DR3 database. It has a parallax that differs 
from that of M10 at roughly the $1\sigma$ level, and so 
it may be in the field. The other variable, 5533, has a parallax that indicates 
it is a foreground object.

	The photometric properties of stars that are flagged 
as variable in GAIA DR3 have also been examined. Of the 12 stars considered, 
we find four that have amplitude fluctuations that are near the upper 
envelope of the plate-to-plate dispersions of stars of the 
same brightness. This does not mean that the other eight stars are not 
variable. Rather, the variations in their light levels are 
too small to register in these data.

	Of the four GAIA variables with larger $\sigma$s, two are 
foreground objects, and the variability type can not be identified from the 
plate photometry. The other two have parallaxes that indicate 
they are at the same distance as M10. One shows light variations 
that are suggestive of a SR, while the other has 
photometric properties that are suggestive of an RR Lyrae star. 
The suspected SR variable appears to be an outlying cluster member. 
The suspected RR Lyrae variable falls above the M10 HB on the cluster CMD, 
suggesting that it is evolving on the supra-horizontal 
branch if it is a member of M10.

	Four other GAIA variables have photometric 
properties that place them squarely on the M10 HB, 
and any variability in these stars is likely less than the 10\% level 
given the plate-to-plate dispersions in their photometry.
Despite having distances that are comparable to that of M10, 
these stars have projected offsets from the cluster center of many parsecs, 
bringing cluster membership into question. A fifth low amplitude GAIA 
variable is located close to the cluster center, and has photometric 
properties that are consistent with evolution on the supra-horizontal branch.

	In closing, this paper has demonstrated that plates 
in the DAO collection that are digitized with a commercial scanner
and that have cosmetic characteristics that are far from ideal 
can yield photometric measurements with a plate-to-plate consistency of 
ten percent or less. This is achieved after applying 
processing that includes the removal of scattered light 
and the balancing of the photometric zeropoint across the plates.
While extracting photometric measurements from plates of this age is a 
non-trivial task, the potential rewards are significant for 
studies of the long-term behaviour of objects. It is hoped that the 
lessons learned here may prove useful to the examination of similar 
older photographic plates by others. 

\acknowledgements{It is a pleasure to thank the anonymous reviewer for 
comments that greatly improved the paper. This 
research has made use of the NASA/IPAC Infrared 
Science Archive (https://doi.org/10.26131/irsa1), which is funded by the
National Aeronautics and Space Administration and operated
by the California Institute of Technology. This work has also
made use of data from the European Space Agency (ESA)
mission Gaia (https://www.cosmos.esa.int/gaia), processed
by the Gaia Data Processing and Analysis Consortium (DPAC,
https://www.cosmos.esa.int/web/gaia/dpac/consortium).
Funding for the DPAC has been provided by national
institutions, in particular the institutions participating in the
Gaia Multilateral Agreement.}

\appendix

	The calibrated photometric measurements of the variables discussed 
in this paper are presented in Table 5. The calibration was 
done using the relations discussed in Section 4.2. The variable names 
that do not have a 'V' designation are the numbers assigned by DAOPHOT, and 
the on-sky locations of those stars can be found in Tables 3 and 4. 
The 'G' preceeding a number indicates that the star 
has been flagged as a variable in the GAIA DR3.
The calibrated measurements for 3744 should be viewed with caution 
because of its very red color, although plate-to-plate 
magnitude differences will be more robust.

\begin{deluxetable}{lccccccccccc}
\tablecaption{$V$ Photometric Measurements}
\tablehead{JD & V2 & V3 & V22 & V24 & V27 & V29 & G3744\tablenotemark{a} & 5533 & G6071 & G2262 & G4003 \\
 -2420000 & & & & & & & & & & & \\}
\startdata
6915.796 & 11.650 & 12.537 & 14.580 & 13.929 & 11.663 & 11.846 & 11.435 & 11.580 & 13.773 & 14.448 & 14.325 \\
6921.742 & 12.272 & 12.329 & 14.604 & 14.064 & 11.941 & 11.931 & 11.649 & 11.935 & 13.863 & 14.315 & 14.371 \\
6923.804 & 11.706 & 12.184 & 14.637 & 14.646 & 11.878 & 11.911 & 12.093 & 11.444 & 13.982 & 14.493 & 14.142 \\
6924.739 & 11.661 & 12.420 & 14.417 & 13.572 & 11.675 & 11.828 & 11.695 & 11.533 & 13.842 & 14.443 & 14.255 \\
6944.735 & 11.811 & 12.199 & 14.459 & 14.107 & 11.836 & 12.057 & 12.119 & 12.070 & 13.946 & 14.630 & 14.263 \\
6944.744 & 11.533 & 12.059 & 14.473 & 13.866 & 11.740 & 11.968 & 12.234 & 11.791 & 13.699 & 14.722 & 14.201 \\
6946.713 & 11.843 & 12.476 & 14.310 & 14.047 & 11.829 & 12.112 & 12.587 & 12.067 & 13.912 & 14.425 & 14.303 \\
6946.722 & 11.550 & 12.290 & 14.375 & 14.058 & 11.810 & 11.990 & 11.729 & 11.761 & 13.841 & 14.416 & 14.526 \\
7274.752 & 12.802 & 12.807 & 14.672 & 14.213 & 12.108 & 12.303 & 11.054 & 12.648 & 14.018 & 14.479 & 14.444 \\
7275.751 & 12.772 & 12.290 & 14.487 & 15.978 & 12.542 & 12.486 & 11.217 & 12.675 & 14.364 & 14.474 & 14.326 \\
7307.754 & 12.172 & 12.627 & 14.258 & 14.505 & 12.183 & 12.361 & 11.986 & 12.323 & 14.031 & 14.660 & 14.208 \\
7308.712 & 11.772 & 12.341 & 14.419 & 14.209 & 11.879 & 12.023 & 12.301 & 11.849 & 13.801 & 14.753 & 14.076 \\
7309.695 & 12.128 & 12.521 & 14.573 & 13.984 & 12.181 & 12.246 & 12.384 & 12.483 & 14.172 & 14.463 & 14.341 \\
7658.724 & 12.546 & 13.007 & 14.537 & 14.327 & 12.181 & 12.389 & 11.196 & 12.642 & 14.395 & 14.310 & 14.334 \\
7659.776 & 11.231 & 12.197 & 14.598 & 13.960 & 11.762 & 12.091 & 11.510 & 12.037 & 13.770 & 14.577 & 14.060 \\
7664.783 & 11.701 & 13.302 & 14.320 & 13.782 & 11.580 & 12.310 & 11.643 & 11.914 & 13.857 & 14.596 & 14.088 \\
\enddata
\tablenotetext{a}{The transformed measurements of this star are subject 
to uncertainties given its very red color.}
\end{deluxetable}

\end{document}